\documentclass[twocolumn]{aa} 

\usepackage{graphicx}
\usepackage{txfonts}
\usepackage{hyperref}          
\hypersetup{
 colorlinks=true,  
 urlcolor=blue,    
 linkcolor=blue,
 citecolor=blue 
}

\usepackage{subfigure}
\usepackage{natbib}
\begin{document}

   \title{Accelerated particle beams in a 3D simulation of the quiet Sun}
   \subtitle{Lower atmospheric spectral diagnostics}
   

   \author{H.~Bakke\inst{\ref{inst1}, \ref{inst2}}
          \and
          L.~Frogner\inst{\ref{inst1}, \ref{inst2}}
          \and
          L.~Rouppe van der Voort\inst{\ref{inst1}, \ref{inst2}}
          \and
          B.~V.~Gudiksen\inst{\ref{inst1}, \ref{inst2}}
          \and
          M.~Carlsson\inst{\ref{inst1}, \ref{inst2}}
          }

   \institute{
   Institute of Theoretical Astrophysics, 
   University of Oslo,
   P.O.Box 1029 Blindern, 
   N-0315 Oslo,
   Norway \label{inst1}
   \and
   Rosseland Centre for Solar physics,
   University of Oslo, 
   P.O.Box 1029 Blindern, 
   N-0315 Oslo,
   Norway \label{inst2}
   }

   \date{}

 
  \abstract
   {Nanoflare heating through small-scale magnetic reconnection events is one of the prime candidates to explain heating of the solar corona. However, direct signatures of nanoflares are difficult to determine, and unambiguous observational evidence is still lacking. Numerical models that include accelerated electrons and can reproduce flaring conditions are essential in understanding how low-energetic events act as a heating mechanism of the corona, and how such events are able to produce signatures in the spectral lines that can be detected through observations.}
   {We investigate the effects of accelerated electrons in synthetic spectra from a 3D radiative magnetohydrodynamics simulation to better understand small-scale heating events and their impact on the solar atmosphere.}
   {We synthesised the chromospheric \ion{Ca}{ii} and \ion{Mg}{ii} lines and the transition region \ion{Si}{iv} resonance lines from a quiet Sun numerical simulation that includes accelerated electrons. We calculated the contribution function to the intensity to better understand how the lines are formed, and what factors are contributing to the detailed shape of the spectral profiles.}
   {The synthetic spectra are highly affected by variations in temperature and vertical velocity. Beam heating exceeds conductive heating at the heights where the spectral lines form, indicating that the electrons should contribute to the heating of the lower atmosphere and hence affect the line profiles. However, we find that it is difficult to determine specific signatures from the non-thermal electrons due to the complexity of the atmospheric response to the heating in combination with the relatively low energy output ($\sim 10^{21}$~erg~s$^{-1}$). Even so, our results contribute to understanding small-scale heating events in the solar atmosphere, and give further guidance to future observations.}
   {}


   \maketitle
%

\section{Introduction}

Nanoflares are heating events associated with small-scale magnetic reconnection in the solar atmosphere. They release energy in the range $10^{24}$--$10^{25}$~erg, and they are believed to occur frequently throughout the atmosphere. The nanoflare heating mechanism is one of the prime candidates in understanding why the corona is heated to millions of Kelvin \citep{1988ApJ...330..474P}. 
It is generally accepted that flare energy is transported by electrons accelerated to non-thermal energies as magnetic field lines reconnect. The accelerated electrons transfer energy to the ambient plasma through Coloumb collisions as they travel along the magnetic field \citep{1971SoPh...18..489B, 1978ApJ...224..241E, 2011SSRv..159..107H}, 
leaving observable signatures in the spectral lines that form in the sites where the energy is deposited. Signatures of non-thermal electrons are found in observed hard X-ray spectra from active region flares. However, X-ray observations of small-scale events with nanoflare energies are rare because the signatures are typically below the detection threshold \citep[although, see e.g.][]{2017ApJ...844..132W, 2020ApJ...891L..34G, 2021MNRAS.507.3936C}. 
As a result, the presence and properties of nanoflares in the solar atmosphere remain poorly known. 

Heating signatures from energetic events in the corona are difficult to observe directly, as the high conductivity of coronal plasma has a tendency to smear the signatures out. It is therefore beneficial to look for signatures of heating release in the atmospheric layers that are responsive to heating, such as the transition region (TR) and chromosphere. Non-thermal electrons accelerated by magnetic reconnection in the corona collide with the dense TR and chromospheric plasma, giving rise to changes in temperature and density. However, looking for specific signatures is problematic as nanoflares are difficult to observe. Through numerical simulations, \cite{2014Sci...346B.315T} 
have found that non-thermal electrons are necessary to reproduce blueshifts in the \ion{Si}{IV}~140.3~nm line observed with the Interface Region Imaging Spectrograph \citep[IRIS;][]{2014SoPh..289.2733D} 
in small heating events at the footpoints of transient hot loops. By exploring a wide range of parameters, \cite{2018ApJ...856..178P} 
have carried out an extensive numerical investigation to better understand and interpret TR observations, and \cite{2022A&A...659A.186B} 
have extended the analysis to include spectral lines that form deeper in the atmosphere and are readily accessible by ground-based telescopes. In the latter, the analysis of chromospheric spectra in 1D simulations of nanoflares showed that the lines forming deeper in the chromosphere experience similar effects as the lines forming higher up.
\citet{2020ApJ...889..124T} 
have further demonstrated that observations of high variability ($\lesssim 60$~s) at the footpoints of hot coronal loops ($\sim$ 8--10~MK) in active region (AR) cores provide powerful diagnostics of the properties of coronal heating and energy transport when combined with numerical simulations.  

In this work, we investigate the effect of accelerated electrons in a 3D radiative magnetohydrodynamics (MHD) simulation by analysing synthetic chromospheric \ion{Ca}{ii}~854.2~nm, \ion{Ca}{ii}~H and K, and \ion{Mg}{ii}~h and k spectral lines as well as the TR \ion{Si}{iv} resonance lines. The simulation is based on the 3D MHD Bifrost model introduced in \cite{2018A&A...620L...5B}, 
which has been further developed in \cite{2020A&A...643A..27F} 
and \cite{2022arXiv221001609F} 
to include a more accurate method for calculating the electron beam heating. We explore the impact of non-thermal electrons by comparing the spectral line analysis from different regions that are both subject and not subject to beam heating. In the analysis we investigate the Doppler shifts of the spectra and the formation of line intensity.

\section{Method}

\subsection{Bifrost simulation}

The numerical simulation was performed using the Bifrost code \citep{2011A&A...531A.154G}, 
which solves the resistive MHD partial differential equations in 3D, with radiative transfer and field-aligned thermal conduction accounted for in the energy equation. In the photosphere and lower chromosphere, Bifrost solves the optically thick radiative transfer equation with scattering \citep{2010A&A...517A..49H}. 
In the upper chromosphere and TR, it approximates non-LTE radiative losses based on parameterised results from 1D radiative hydrodynamic simulations \citep{2012A&A...539A..39C}. 
Finally, it calculates optically thin radiative losses in the corona.

Bifrost was developed with a high degree of modularity, allowing users to extend the code with additional physics. In \citet{2018A&A...620L...5B}, 
we presented a method for treating energy transport by accelerated electrons in Bifrost simulations, which we expanded upon and discussed in depth in \citet{2020A&A...643A..27F} 
and \citet{2022arXiv221001609F}. 
The first step of the method is to detect locations where the magnetic field reconnects using a criterion for reconnection in MHD theory \citep{2005mrp..book.....B}. 
The second step is to estimate the energy distributions of the non-thermal electrons expected to be accelerated at each reconnection site. We assume that the distribution is a power-law, with a lower cut-off energy $E_\mathrm{c}$ corresponding to the intersection of the power-law with the local thermal distribution. The lower cut-off energy is not fixed, but is roughly proportional to temperature (for example, $10^6$~K corresponds to a lower cut-off energy of the order of 1~keV). We determine the total non-thermal energy flux assuming that a fixed fraction $p$ of the released magnetic energy, which otherwise would be converted entirely into resistive heating, goes into accelerating electrons. The value of $p = 0.2$ was chosen based on flare observations suggesting that typical values of $p$ range from 10\% \citep{2004JGRA..10910104E, 2012ApJ...759...71E} 
up to 50\% \citep{1971SoPh...17..412L}. 
Finally, we leave the power-law index $\delta$ as a free global parameter. This parameter largely affects the resulting distribution of deposited electron beam energy. We used a value of $\delta = 6$, which has a faster rate of deposited energy compared to smaller values. We note that a larger $\delta$ leads to a smaller penetration depth of the beam \citep{2015TESS....130207A}. 
The value of $\delta$ is supported by observational evidence showing an increase in power-law index with decreasing flare energy \citep{2011SSRv..159..263H}. 
A higher value of $\delta$ is also motivated by the 1D flare simulations analysed in \cite{2022A&A...659A.186B}, 
where $\delta = 7$ was used for the non-thermal electron energy distribution.
The final step is to trace the trajectory of each non-thermal electron beam along the magnetic field while computing the heating of the local plasma due to Coulomb collisions along the way. For this, we use an analytical expression accounting for the systematic velocity change of beam particles due to collisions with ambient hydrogen atoms and free electrons \citep{1978ApJ...224..241E, 1994ApJ...426..387H}. 
During the simulation, we continually computed the transfer of energy by the beams in this way and included it as a term in the MHD energy equation.

\begin{figure}[!t]
    \includegraphics[width=\columnwidth]{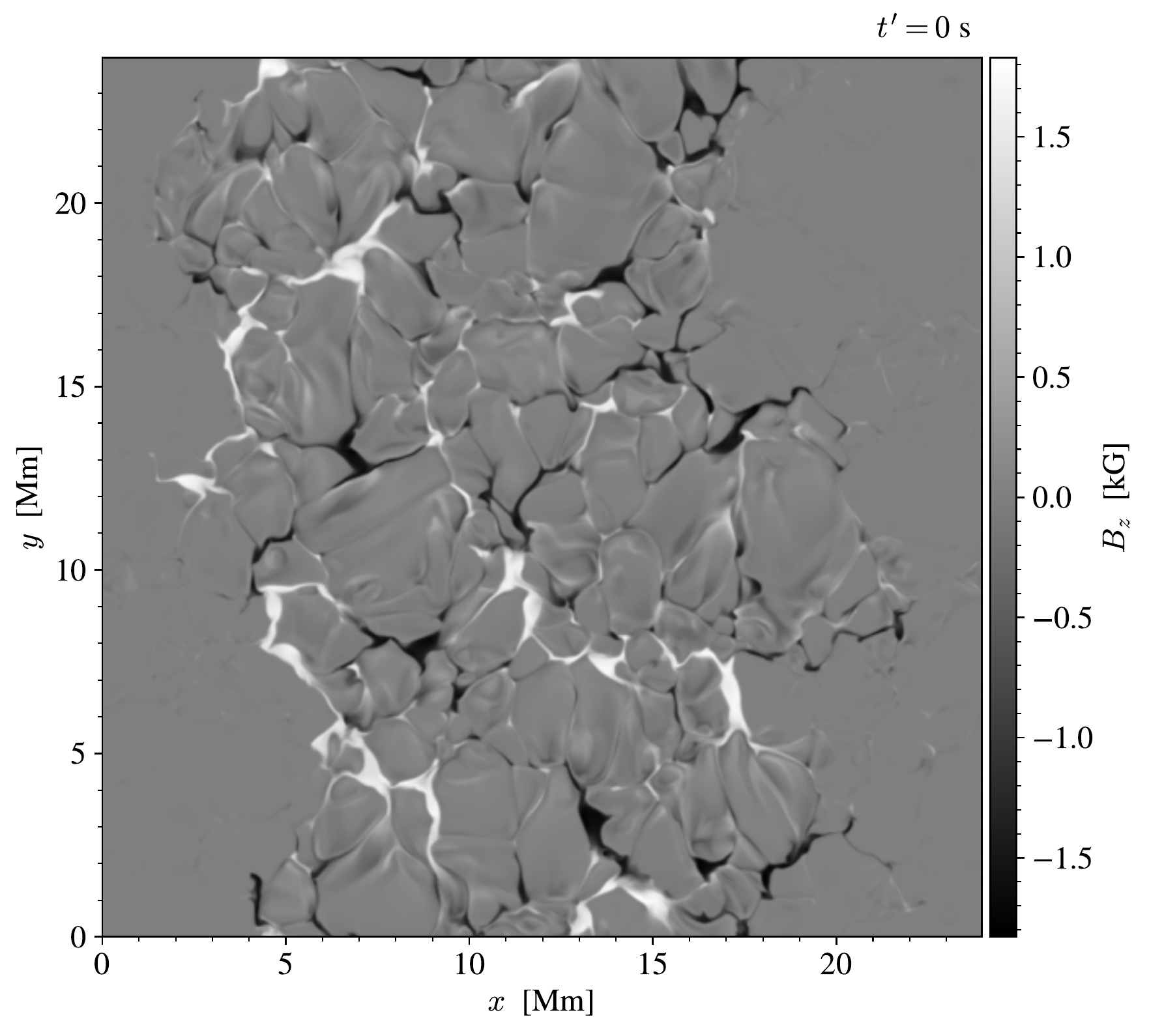}
    \centering
    \caption{Vertical magnetic field $B_z$ in the photosphere at $t^\prime = 0$~s (8220~s after the magnetic flux sheet has been injected).}
    \label{fig:mag_field}
\end{figure}

The particular atmospheric simulation considered in this paper encompasses a horizontal area of $24 \times 24$~Mm and a vertical span from 2.5~Mm below the photosphere to 14.3 Mm above it, in the corona. The simulation has a resolution of $768 \times 768 \times 768$ grid cells, with a uniform grid cell extent of 31~km in the horizontal directions and uneven vertical grid cell extents that vary with height in the atmosphere. Due to the need to resolve sudden local variations near the transition region, the grid cells are about 12~km tall between the photosphere and the height of 4~Mm. From this region, the vertical grid cell extent increases evenly to 21~km at the bottom of the simulation box and to 80~km at the top. In the simulation, heating at the bottom boundary in combination with radiative cooling in the photosphere produces convective motions. The chromosphere and corona are heated by magnetic reconnection and acoustic shocks resulting from these motions. At this point, the Bifrost simulation qualifies as rather quiet, and electron beams from this level of reconnection can be regarded as weak. In order to perturb the system with more magnetic energy and produce more energetic reconnection events, we introduced a large scale magnetic flux emergence. 
To emulate flux emergence, a sheet with magnetic field strength of 2000~G oriented in the $y$-direction was injected at the bottom boundary. As it rose up through the convection zone and coalesced in the convective downflow regions, the injected field organised into a largely bipolar loop system pushing up on the ambient $x$-directed magnetic field that was originally present in the corona. Reconnection between the ambient and injected field then resulted in minor energy release and particle acceleration events throughout the corona. This setup made the simulation more active, but still relatively quiet as compared to solar active regions with high flaring activity. We note that the original setup of this Bifrost simulation was developed and used by \cite{2019A&A...626A..33H}, 
with the aim of studying the generation of Ellerman bombs and UV bursts through flux emergence.

For this paper, we consider a series of 36 snapshots with 1~s intervals, where the simulation time step is $10^{-3}$~s. This simulation starts 8220~s after the magnetic flux sheet has been injected at the bottom boundary. The vertical component of the magnetic field in the photosphere at this time can be seen in Fig.~\ref{fig:mag_field}, where $t^\prime = 0$~s is the first time step where the electrons are injected. 
The total power of accelerated electrons in a single simulation snapshot is roughly $10^{24}$~erg~s$^{-1}$, and individual beams along the magnetic field produce approximately $3 \cdot 10^{21}$~erg~s$^{-1}$ of non-thermal power \citep{2020A&A...643A..27F}. 
A typical small-scale beam heating event is estimated to release $10^{20}$--$10^{24}$~erg of non-thermal energy in the lower atmosphere assuming that the event lasts around 100~s. With 36~s of simulation time, we can assume that the energy released by heating events is on the lower end of this range, and hence a few orders of magnitudes less than the typical nanoflare energy (which is about $10^{24}$--$10^{25}$~erg). But even though the events are weak, they are highly abundant, and a significant number of small beam heating events are likely to occur in the chromosphere at any given time. 
We note that it is difficult to provide a meaningful number of events as they have a tendency to lose their identity in the simulation due to the relatively low energy released. However, see \cite{2017A&A...603A..83K}  
for a method of identifying coronal heating events in a Bifrost simulation by detecting 3D volumes of high Joule heating to find locations with current sheets.

\subsection{Spectral synthesis with RH}
The spectral lines were synthesised using the RH1.5D radiative transfer code \citep{2001ApJ...557..389U, 2015A&A...574A...3P}, 
which calculates spectra from 1D, 2D or 3D numerical simulations on a column-by-column basis. RH1.5D solves the non-LTE radiative transport for spectral lines in partial redistribution (PRD), which is important in the synthesis of chromospheric lines where a more accurate treatment of photon scattering is required. While PRD is not strictly necessary in the synthesis of all our chosen spectra, RH1.5D can still be employed as a general non-LTE code. In general, PRD is assumed in the synthesis of \ion{Mg}{ii}~h and k \citep{1974ApJ...192..769M, 2013ApJ...772...89L, 2013ApJ...772...90L} 
and \ion{Ca}{ii}~H and K \citep{1974SoPh...38..367V, 1975ApJ...199..724S, 2018A&A...611A..62B}, 
but is less important for \ion{Ca}{ii}~854.2~nm and the \ion{Si}{iv} resonance lines. 

Each atmosphere from the Bifrost snapshots was used as input to the RH code. We did not include a micro-turbulence term to the spectral synthesis as we wanted to focus on the effect from velocities in the Bifrost simulation. The $z$-axis of the input atmospheres includes the heights from the surface to the corona, excluding the convection zone as it is not relevant for the line synthesis.
We selected all columns in $x$- and $y$-direction when synthesising the spectra from the main snapshot at $t^\prime = 28$~s analysed in this study, covering a domain of $768 \times 768 \times 670$ pixels. The synthetic spectra for the entire time series were calculated using a coarser grid ($384 \times 384 \times 670$ pixels) for the model atmospheres in order to reduce the computation time. The coarser grid does not affect the spectral analysis as the level of change from one grid cell to its neighbouring ones is almost negligible. A similar coarser sampling
of a Bifrost simulation domain was performed in \cite{2013ApJ...772...90L}. 
We used the default 5 level-plus-continuum \ion{H}{i} and \ion{Ca}{ii} atoms, the 10 level-plus-continuum \ion{Mg}{ii} atom from \cite{2013ApJ...772...89L}, 
and the 30 level-plus-continuum \ion{Si}{iv} atom from \cite{2019ApJ...871...23K}. 
The latter was used to allow the \ion{Si}{iv} resonance lines to form under optically thick conditions, as the model silicon atom includes potential opacity effects. It is common to assume that the \ion{Si}{iv} emission is formed under optically thin conditions, and hence compute the emissivity without calculating the full radiative transfer. Through 1D flare modelling, \cite{2019ApJ...871...23K} 
found that optical depth effects are considerable in the produced \ion{Si}{iv} emission, and that the lines can form under optically thick conditions even for weaker flares with electron energy flux down to $F \approx 5 \cdot 10^9$~erg~cm$^{-2}$~s$^{-1}$. 
We note that the model atom was constructed for use on simulated flares in RADYN \citep{1992ApJ...397L..59C, 1995ApJ...440L..29C, 1997ApJ...481..500C, 2015TESS....130207A}, 
a 1D radiative transfer code that allows for flare investigation in an isolated system. The model atom employs a photospheric value $A_\mathrm{Si} = 7.51$ for the silicon abundance \citep{2009ARA&A..47..481A}, 
even though other work \citep[e.g.][]{2015ApJ...802....5O, 2016ApJ...817...46M} 
have argued in favour of using coronal abundances for silicon and other low first ionisation potential (low-FIP) elements. Using coronal abundances is based on the findings that low-FIP elements tend to be overabundant in the TR and corona \citep{2004ApJ...614.1063L}. 
However, \citet{2016ApJ...824...56W} 
have shown that low-FIP elements have a composition that is close to that of the photosphere during impulsive heating events. In Bifrost, it is possible that the model silicon atom is more accurate in regions that are subject to electron acceleration, but also at the sites where the electron energy is deposited. However, we keep in mind that the silicon abundance might not be accurate in areas that are not subject to heating. 

The synthetic spectra were calculated for 36~s, corresponding to 36 Bifrost snapshots with a 1~s time interval. IRIS observations of TR moss show that the lifetime of short-lived brightenings resulting from coronal nanoflare heating at the footpoints of hot transient coronal loops varies between 10--30~s \citep{2013ApJ...770L...1T, 2014Sci...346B.315T, 2020ApJ...889..124T}. 
With the Bifrost time series, we should be able to study the effects of non-thermal electrons on a similar timescale. However, we note that with the current energy distribution for the non-thermal electrons, it is difficult to obtain a strong signal in the synthetic spectra. 

\subsection{Optically thin calculation of \ion{Si}{iv} emission}

While the RH spectral line synthesis allows for the \ion{Si}{iv} resonance lines to form under both optically thick and thin conditions, we also include a more straight forward approach to calculate the \ion{Si}{iv} emission under optically thin conditions in order to compare potential differences and put further constraints on the interpretation of observations. The approach we used is similar to that of \citet{2015ApJ...802....5O}, 
where we calculated emissivities for the relevant \ion{Si}{iv} energy transitions using atomic data from the CHIANTI database \citep{1997A&AS..125..149D, 2021ApJ...909...38D}. 
We did this for a range of temperatures and electron densities representative of the conditions in the corona and upper TR of the simulated atmosphere to create a lookup table enabling us to efficiently obtain the emissivity at every location in the simulation domain. We note that emissivities for temperatures lower than 10\,000~K 
are set to zero in CHIANTI. In this approach, we used the coronal abundance $A_\mathrm{Si} = 8.10$ \citep{1992PhyS...46..202F}. 
To determine the intensities formed in the optically thin regime as it emerges from the atmosphere, we integrated the emissivities in each vertical column of the atmosphere. We also computed the Doppler shift and width of the synthetic spectral line by evaluating the first and second moment of the locally emitted line profile with respect to Doppler shift from the line centre, and integrated this over height. We assume that the locally emitted radiation has a Gaussian spectral profile with thermal broadening and a Doppler shift depending on the local plasma velocity. The optically thin calculation is also significantly cheaper in computational terms than the RH spectral synthesis. 

\begin{figure}[!ht]
    \centering
    \includegraphics[width=\columnwidth]{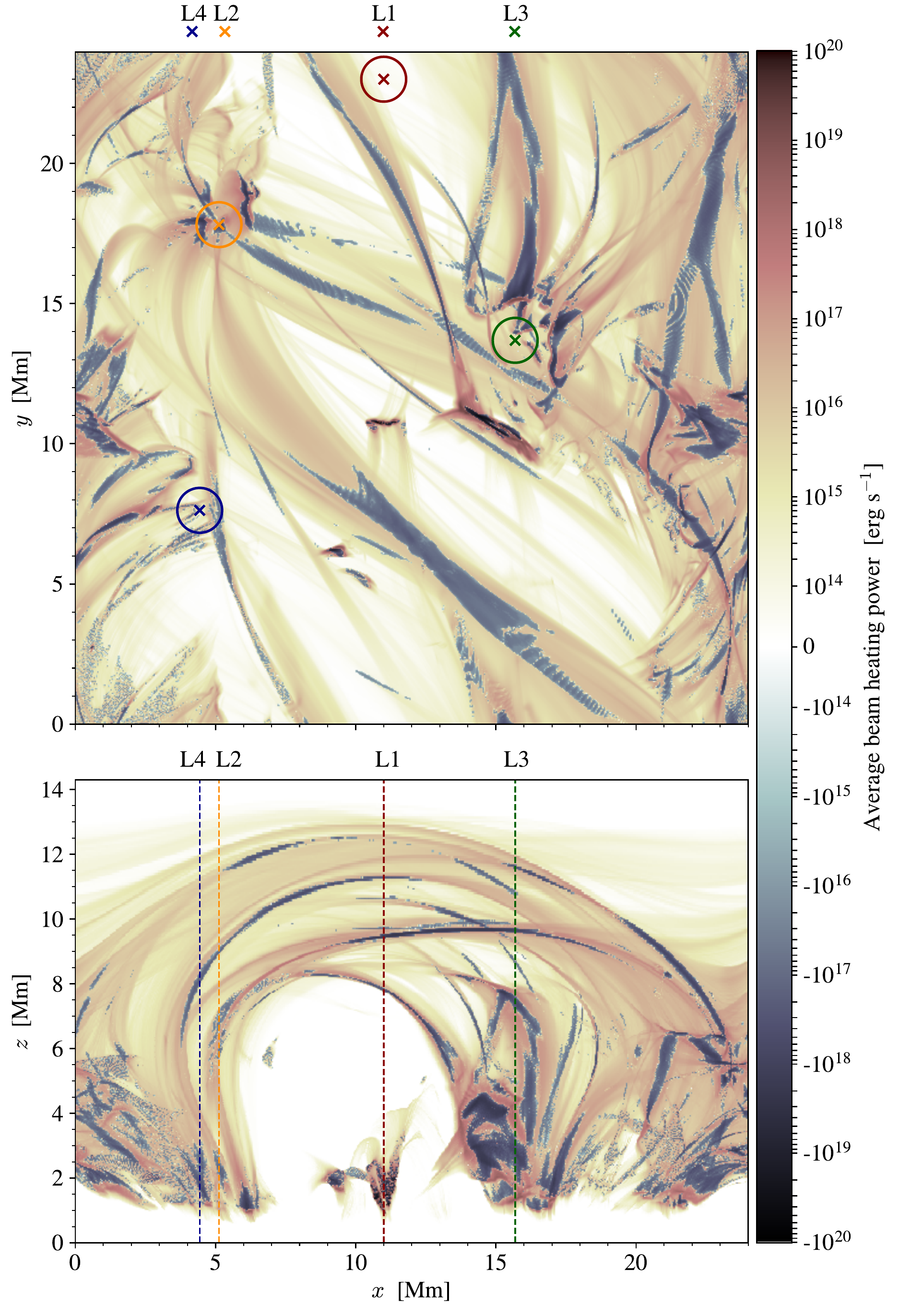}
    \caption{Integrated average beam heating power along the $z$-axis (upper panel) and $y$-axis (lower panel) of the Bifrost simulation snapshot at $t^\prime = 28$~s. The blue regions represent where a fraction $p$ of the reconnection energy is being injected into the accelerated electrons, while the orange regions show where the non-thermal electrons deposit their energy and heat the plasma. The coloured circles in the upper panel are the areas of interest, while the crosses (upper panel) and dashed lines (lower panel) mark the specific locations analysed in detail (L1, L2, L3, and L4).}
    \label{fig:sel_areas}
\end{figure}

\subsection{Contribution function to the line intensity} 

The spectral diagnostics consisted of analysing the contribution function to the emergent intensity. The contribution function can be used to explore which parts of the atmosphere contributes to the line formation. Following \citet{1997ApJ...481..500C}, 
the contribution function was calculated as
\begin{equation} \label{eq:cont_func}
    C_{I_\nu}(z) \equiv \frac{\mathrm{d}I_\nu(z)}{\mathrm{d}z} = S_\nu \ \tau_\nu \mathrm{e}^{-\tau_\nu} \ \frac{\chi_\nu}{\tau_\nu}.  
\end{equation}
The first term on the right-hand side gives the total source function $S_\nu$. Here, $S_\nu$ is dependent on frequency because we assume PRD. The next term is the optical depth factor $\tau_\nu \mathrm{e}^{-\tau_\nu}$, which represent the Eddington-Barbier part of the contribution function. The optical depth factor has a maximum at $\tau_\nu = 1$. The final term, $\chi_\nu/\tau_\nu$, is the ratio of the opacity over optical depth. The term is responsible for line asymmetries due to its sensitivity to velocity gradients in the atmosphere. In the presence of strong velocity gradients, the opacity is typically large at small optical depths, and $\chi_\nu/\tau_\nu$ is the dominant factor in the contribution function. 

\subsection{Locations of interest}

The locations of interest were selected based on the electron acceleration regions. Figure~\ref{fig:sel_areas} shows the net electron beam heating power integrated vertically (upper panel) and horizontally (lower panel) over the simulation domain. The electrons are mostly accelerated along the magnetic field, where negative average beam heating power (blue regions) indicates where part of the energy is transported away from the reconnection site. As shown in the upper panel, we have chosen three areas (orange, green, and blue circles) at magnetic field footpoints that are associated with field lines were electrons are accelerated, and a reference area (red circle) without beam impact. The crosses (upper panel) and dashed lines (lower panel) represent the specific locations L1, L2, L3, and L4 that we subsequently analyse in detail.

The specific locations were found using CRISPEX \citep{2012ApJ...750...22V}, 
which is a widget based tool developed to browse and analyse large observational data sets. However, CRISPEX can also be used to analyse synthetic spectra from simulations by creating a data cube that is readable by the tool. We formatted the synthetic \ion{Mg}{ii}~k, \ion{Ca}{ii}~K, and \ion{Si}{iv}~140.3~nm lines from Bifrost at $t^\prime = 28$~s as a multidimensional data cube that is readable by CRISPEX, and chose locations based on the intensity and complexity of the profiles by browsing through the spectra within the different areas. 
We further note that the prominent low altitude current sheet at $(x, y) = (10, 11)$ was found to produce an Ellerman bomb and UV burst in the detailed analysis by \cite{2019A&A...626A..33H}. 

\section{Results}

\subsection{Evolution of the Bifrost atmosphere}

\begin{figure*}[!ht]
    \centering
    \includegraphics[width=\textwidth]{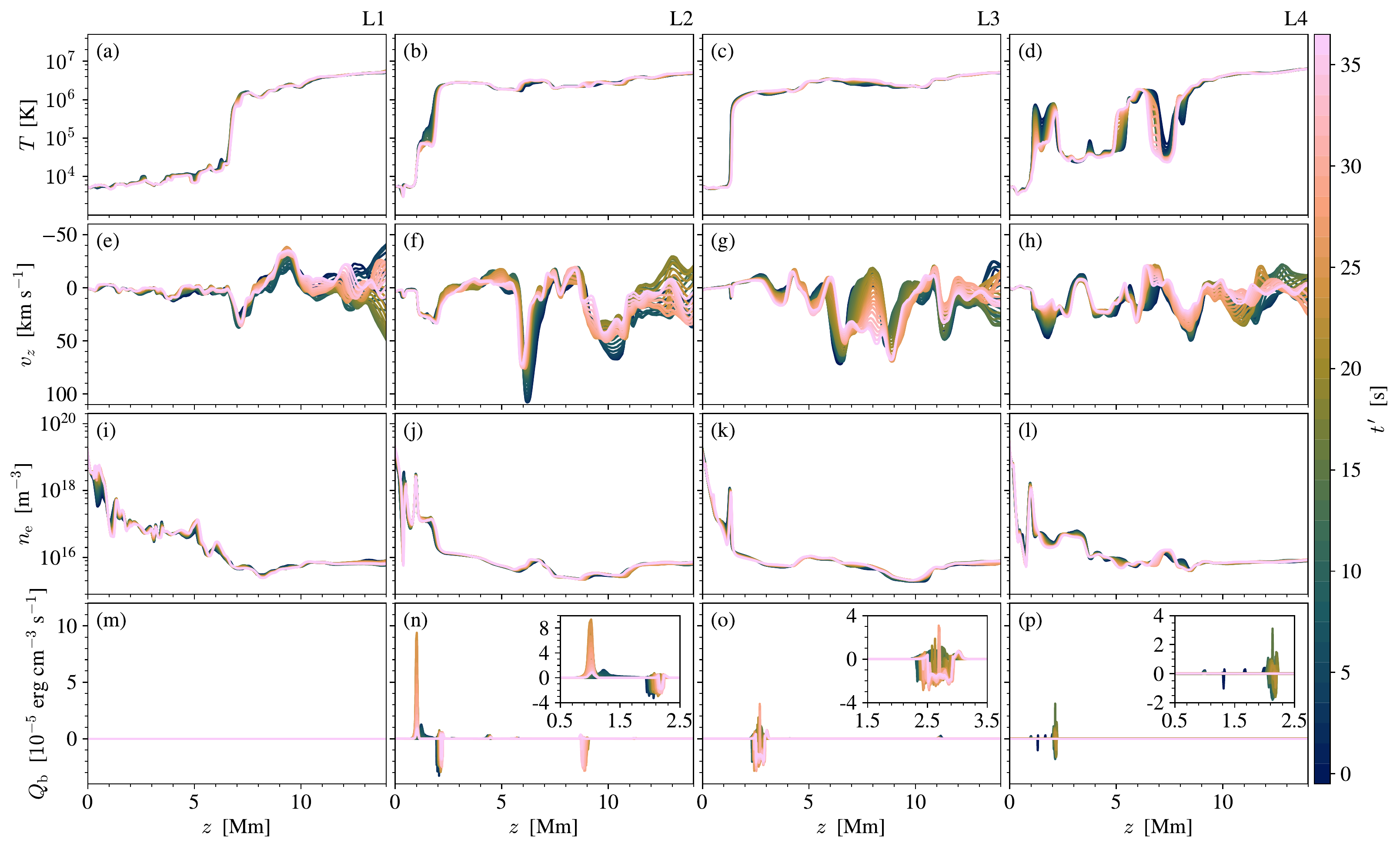}
    \caption{Evolution of temperature $T$, vertical velocity $v_z$, electron number density $n_\mathrm{e}$, and beam heating rate $Q_\mathrm{b}$ in the Bifrost simulation. The quantities are plotted in the range $z \in [0, 14]$~Mm at 1~s intervals for the duration of the time series. Each column represents the specific locations (L1, L2, L3, and L4) from the chosen areas. Negative (positive) velocities correspond to upflows (downflows). The insets in panels (n)--(p) show the electron beam heating in a sub-region, where the $y$-axes are limited to better show the details of the variations in $Q_\mathrm{b}$.}
    \label{fig:atmos_resp}
\end{figure*}

\begin{figure*}[!ht]
    \centering
    \includegraphics[width=\textwidth]{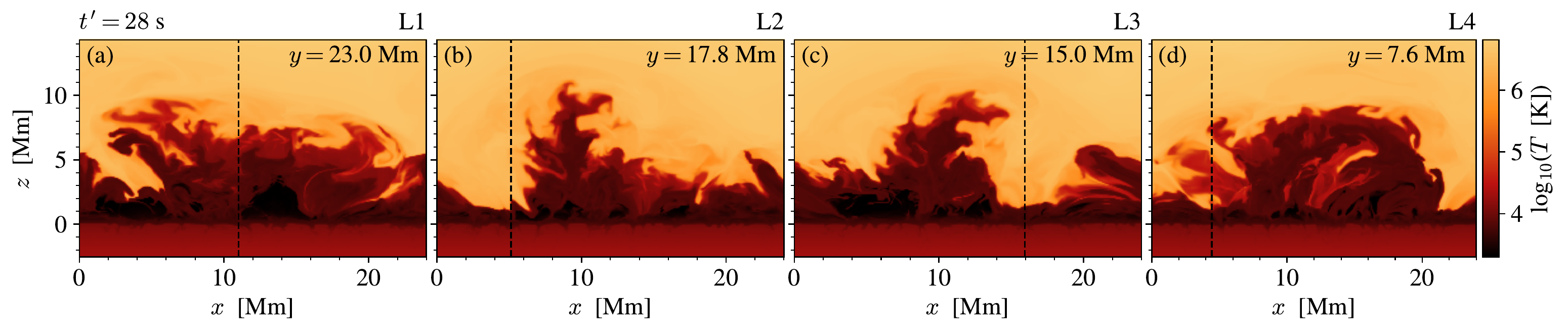}
    \caption{Vertical cut in the $xz$-plane of the temperature structure at $t^\prime = 28$~s. The vertical cut is taken at the location of the $y$-coordinate of L1, L2, L3, and L4. The dashed lines are drawn at the $x$-coordinate of the different locations.}
    \label{fig:temp_cut}
\end{figure*}

Figure~\ref{fig:atmos_resp} shows the time evolution of temperature $T$, vertical velocity $v_z$, and electron number density $n_\mathrm{e}$ in the Bifrost simulation at 1~s intervals. The rows represent the different quantities, while the columns represent the specific locations shown in Fig.~\ref{fig:sel_areas}. The temperature in the four different panels does not experience significant increases or decreases over time. However, the atmospheric structure varies from location to location. This is better seen in Fig.~\ref{fig:temp_cut}, showing the temperature at a vertical cut in the $xz$-plane taken at the location of the $y$-coordinate of L1, L2, L3, and L4, at a single instance in time ($t^\prime = 28$~s). 

L1 is located within the flux emergence region of the simulation (see Fig.~\ref{fig:temp_cut} (a)). The magnetic bubble does not have a significant temperature rise until the canopy is reached around $z = 6.8$~Mm, which can be seen as the TR in Fig.~\ref{fig:atmos_resp} (a). L1 is also located in a region without beam impact, as can be seen in Fig.~\ref{fig:atmos_resp} (m) where $Q_\mathrm{b} = 0$ over the entire duration of the simulation. L2, L3, and L4 are located at magnetic field footpoints, where the TR and temperature rise to coronal values are located at lower altitude. 
Panel (b) in Fig.~\ref{fig:atmos_resp} shows that the TR is gradually pushed upwards over time. This is happening in the height where electrons are both accelerated and depositing their energy (see panel (n)). However, both the velocity and electron number density are relatively unchanged over the duration of the simulation, making it difficult to conclude that the moving TR is caused by the electrons only. If the latter was the case, we would expect the electrons that deposit their energy at 1~Mm to accelerate the dense plasma, causing upflowing (negative) velocities that move the TR to greater heights. Panel (n) shows that the electrons contribute to the thermodynamics of the TR and chromosphere, but the level of heating available in the simulation combined with the complex dynamics of the Bifrost atmosphere causes the effect from the deposited electron beam energy to blend and smear out. This makes it difficult to recognise an unambiguous footprint left by the electrons.  

The temperature at L3 (panel (c)) does not change significantly over the duration of the simulation. The TR is located around 1.5~Mm, and panels (g) and (k) show that at this height there is a weak plasma upflow and an increase in electron number density. We note that the change over time is so small that the line of the last time step covers small variations. There are generally more visible changes over time in the corona compared to the lower atmosphere. The beam heating rate in panel (o) shows that the electrons are both accelerated and depositing their energy between 2.3--3.1~Mm, which is in the corona. This is because the cut-off energy $E_\mathrm{C}$ at coronal heights is low (around 1~keV), and the electrons loose most of their energy through interactions with the coronal plasma. This leads us to suspect that particular features in the synthetic spectra are not caused by the electrons, as signature are potentially only visible in coronal spectral lines and we focus here on spectral lines formed deeper in the atmosphere. 

The temperature structure at L4 (see panel (d) in Figs.~\ref{fig:atmos_resp} and \ref{fig:temp_cut}) is much more complex compared to the other locations. The cool plasma from the magnetic bubble intersects the column at several different heights, making the atmospheric structure difficult to analyse. Panel (p) shows that the beam heating rate is almost balanced out by the energy transferred to the electrons at the reconnection site. This means that the electrons that get accelerated at approximately 2~Mm deposit their energy right away, and do not contribute to any noticeable heating.

\begin{figure}[!hb]
    \includegraphics[width=\columnwidth]{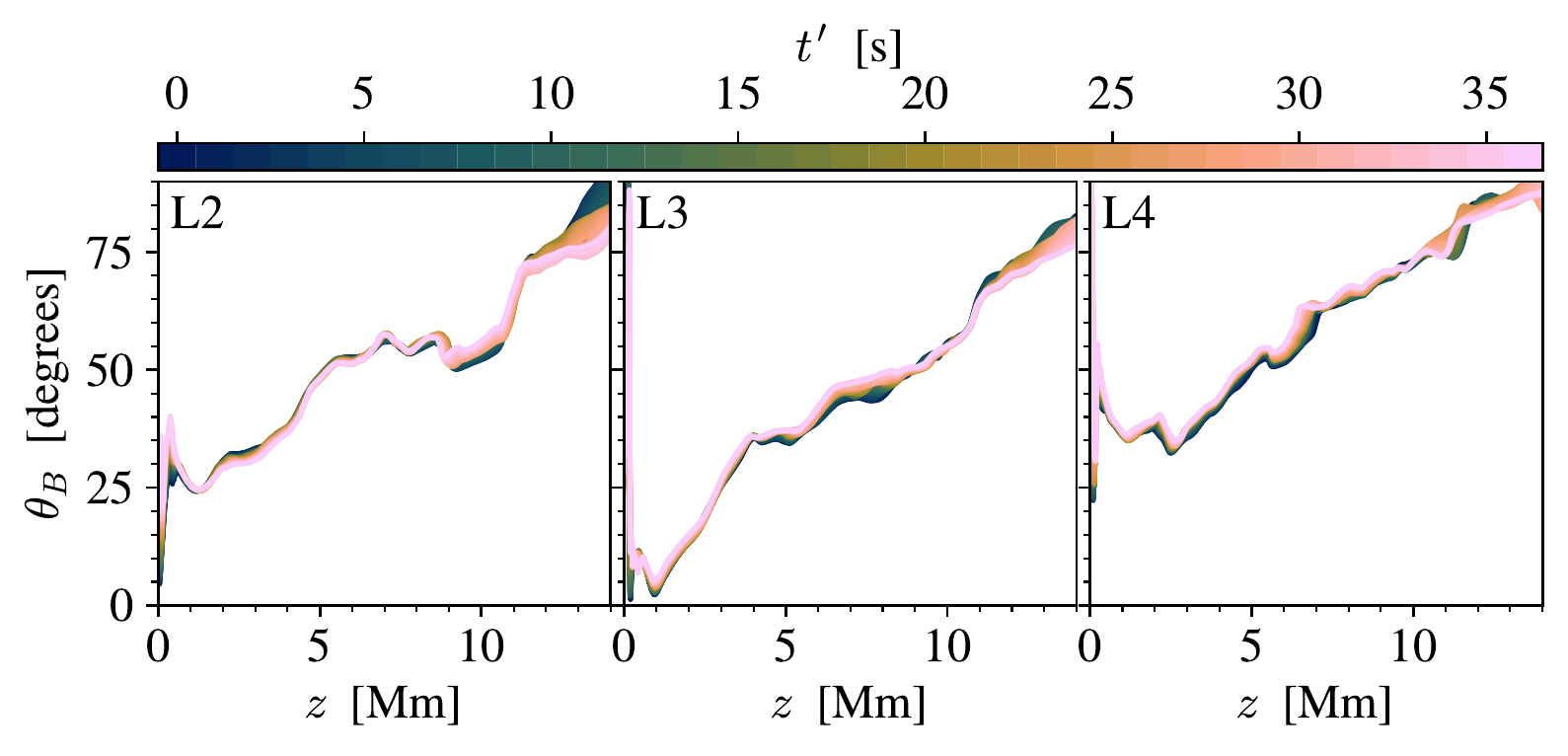}
    \centering
    \caption{Angle between the magnetic field and the vertical direction as a function of $z$ for the entire Bifrost time series at the L2, L3, and L4 locations.}
    \label{fig:field_angle}
\end{figure}

It is important to note that Fig.~\ref{fig:atmos_resp} shows the evolution of the atmosphere along the $z$-axis as seen from directly above, and not along the magnetic field lines. It is beneficial to choose pixel locations that are situated at magnetic field footpoints, as it is possible to detect spectral line signatures from electrons accelerated along the magnetic field connected to the particular footpoint. However, because of the low cross-field transport of energy, we do not expect to see a direct effect of heating by accelerated electrons to travel along $z$ because the effect is isolated to the specific field line where the heating is taking place. This is most likely not along $z$, as the field lines can at best only be regarded as straight at the very bottom of the atmosphere. This is better illustrated in Fig.~\ref{fig:field_angle}, which shows the angle between the field and the vertical direction, calculated as
\begin{equation} \label{eq:field_angle}
    \theta_B \equiv \mathrm{tan}^{-1} \bigg( \frac{\sqrt{B_x^2 + B_y^2}}{|B_z|} \bigg) = \mathrm{tan}^{-1} \bigg( \frac{|B_\mathrm{h}|}{|B_z|} \bigg),
\end{equation}
at the magnetic footpoint locations (L2, L3, and L4) over the duration of the simulation. The figure shows that the angle increases with $z$, reaching $90^\circ$ in the corona where the field lines are mostly horizontal.
We note that at L3 and L4, the angle is $90^\circ$ at $z = 0$~Mm because the magnetic field is highly complex in the convection zone, and the surface is not always located at exactly 0~Mm. This means that the magnetic field might not be aligned with $z$ at 0~Mm, hence we see that the angle between $z$ and the magnetic field is large rather than small. 
We also note that even though the angle is smaller at low heights (for instance around 1~Mm at L3), we are not be able to see direct effects of non-thermal electrons potentially depositing their energy at these heights unless $\theta_B$ is zero. The heating signatures from the electron beams seen in Fig.~\ref{fig:atmos_resp} (n)--(p) do not originate from vertical field lines that are aligned with $z$, but rather from reconnection events along the magnetic field connected to the footpoints. 

\subsection{Emission from synthetic spectra}

\begin{figure*}[!ht]
    \centering
    \includegraphics[width=\textwidth]{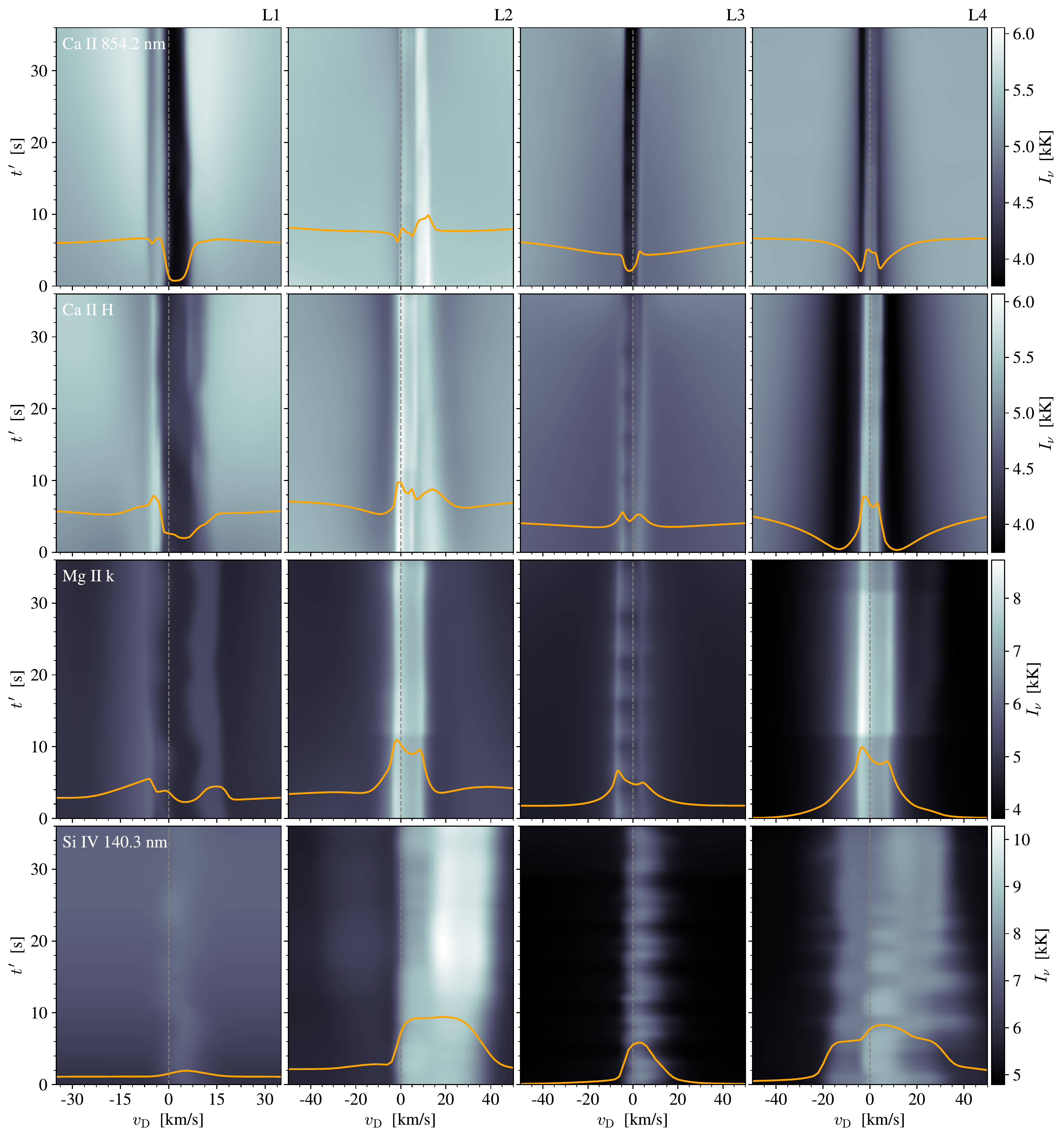}
    \caption{Spectral evolution of \ion{Ca}{ii}~854.2~nm, \ion{Ca}{ii}~H, \ion{Mg}{ii}~k, and \ion{Si}{iv}~140.3~nm at the locations of interest. The $x$-axes are in units of Doppler offset, where negative (positive) velocities indicate blueshifts (redshifts). The intensity is shown in units of brightness temperature. The orange line profiles are taken at $t^\prime = 0$~s, where the highest (lowest) intensity of the profiles in each row corresponds to the maximum (minimum) intensity of the respective colourbars. We note that the intensity of \ion{Si}{iv}~140.3~nm at the L1 location has a maximum of $I_\nu = 7$~kK in order to visually enhance the features of the relatively weak emission. The orange line profiles give a better indication of the difference in intensity across the \ion{Si}{iv} row.}
    \label{fig:multi_int}
\end{figure*}

Figure~\ref{fig:multi_int} represents the time evolution of the synthetic \ion{Ca}{ii}~854.2~nm, \ion{Ca}{ii}~H, \ion{Mg}{ii}~k, and \ion{Si}{iv}~140.3~nm lines at the four selected locations, where we have added individual line profiles at $t^\prime = 0$~s in each panel (orange line) to indicate what the spectra looks like. We note that our findings from the \ion{Ca}{ii}~K, \ion{Mg}{ii}~h, and \ion{Si}{iv}~139.4~nm lines are similar to \ion{Ca}{ii}~H, \ion{Mg}{ii}~k, and \ion{Si}{iv}~140.3~nm, respectively, hence these results are not shown. At L1, the minimum intensity of \ion{Ca}{ii}~854.2~nm at $t^\prime = 28$~s is redshifted to a value between 1 and 2~km~s$^{-1}$. The line seems to narrow towards the end of the simulation, but the general shape of the profile persists. At $t^\prime = 28$~s, the minimum intensity of \ion{Ca}{ii}~H and \ion{Mg}{ii}~k and the single peak of \ion{Si}{iv}~140.3~nm are redshifted to approximately $+5$~km~s$^{-1}$. The \ion{Ca}{ii}~H line profile has increased emission in the blue wing and peak that becomes weaker over time, and there are small variations in the intensity of the line core and the red peak. Over time, the minimum intensity of \ion{Mg}{ii}~k and the single peak of \ion{Si}{iv}~140.3~nm shifts periodically between approximately 0~km~s$^{-1}$ and $+5$~km~s$^{-1}$.

At the L2 location, the synthetic spectra are redshifted to varying degree. The minimum intensity of \ion{Ca}{ii} and \ion{Mg}{ii} are redshifted to a value between 1 and 2~km~s$^{-1}$, while the \ion{Si}{iv} line exhibits a much stronger redshift of the line. The latter is also significantly broadened, most likely due to the downflows at TR heights. \ion{Si}{iv} forms higher in the atmosphere compared to the other spectra, and it is therefore more likely to be affected by the strong downflow seen between 1--2~Mm in Fig.~\ref{fig:atmos_resp} (f). In the beginning of the simulation, there is strong emission in the absorption feature so that the profile almost looks single peaked. As time progresses, the spectral profile becomes broader and the line is redshifted up to approximately $+30$~km~s$^{-1}$. From around 14~s, the red and blue peaks become more pronounced due to increased emission in these components. The \ion{Ca}{ii} lines show increased emission in the red wing that is due to the weak downflowing velocity around 0.4~Mm (see Fig.~\ref{fig:atmos_resp} (f)). Over time, the red wings of the profiles become less broad. This feature is not seen in \ion{Mg}{ii}~k, suggesting that the line forms slightly above this height. 

The evolution of the spectra at L3 shows that the different spectra are experiencing oscillations. This behaviour is not detectable in the \ion{Ca}{ii}~854.2~nm panel, but further investigation show that this line, along with the other spectral lines, are subject to shock waves passing through the atmosphere. Around the formation height of \ion{Ca}{ii}~854.2~nm, we see temperature oscillations varying between 6\,500 and 6\,700~K, but the intensity amplitude is too small to see because of the large range between core minimum and wing maximum. Around the formation height of \ion{Ca}{ii}~H, \ion{Mg}{ii}~k, and \ion{Si}{iv}~140.3~nm, the temperature oscillations vary between 7\,200 and 7\,800~K, 8\,500 and 11\,000~K, and 8\,000 and 14\,000~K, respectively.
These values are low for \ion{Si}{iv}, but we note that the temperatures are taken at the $\tau = 1$ height and that the contribution function covers a wider range.
The temperature oscillations at the formation height of \ion{Si}{iv}~140.3~nm exhibit the largest variation, hence the line is showing the strongest modulation in intensity. As the difference in minimum and maximum temperature decreases, the oscillating pattern in the intensity panels becomes weaker. 

At L4, the \ion{Ca}{ii} spectral profiles have line cores at approximately 0~km~s$^{-1}$ at the beginning of the simulation that are redshifted to a value between 1 and 2~km~s$^{-1}$ over time. Both profiles are double peaked with a slightly more intense blue peak, but the \ion{Ca}{ii}~854.2~nm profile becomes single peaked and less intense as time progresses. The \ion{Ca}{ii}~H profile keeps its double peaked shape, but the red peak becomes less intense from around 20~s. The \ion{Mg}{ii}~k profile is similar to that of L2, with a blue peak that is more intense than the red peak and an absorption feature that is redshifted to a value between 1 and 2~km~s$^{-1}$. At $t^\prime = 12$~s, there is a sudden increase in intensity of the entire line profile that is also faintly seen in the \ion{Mg}{ii}~k panel at the L2 location. This is due to a sudden increase in temperature at the formation height of the spectral line. The \ion{Si}{iv}~140.3~nm line profile is severely broadened over the entire duration of the simulation. The initial profile (orange line) has a red and blue peak and a central reversal of the line core at approximately $+6$~km~s$^{-1}$ that has a higher intensity than the peaks. Similar to the L3 location, the \ion{Si}{iv} line is subject to shock waves, where the temperature oscillations around its formation height vary between 26\,000 and 29\,000~K. At around 26~s, the temperature stabilises and the shape of the profile is similar to the initial profile ($t^\prime = 0$~s).

\begin{figure*}[!ht]
    \centering
    \includegraphics[width=\textwidth]{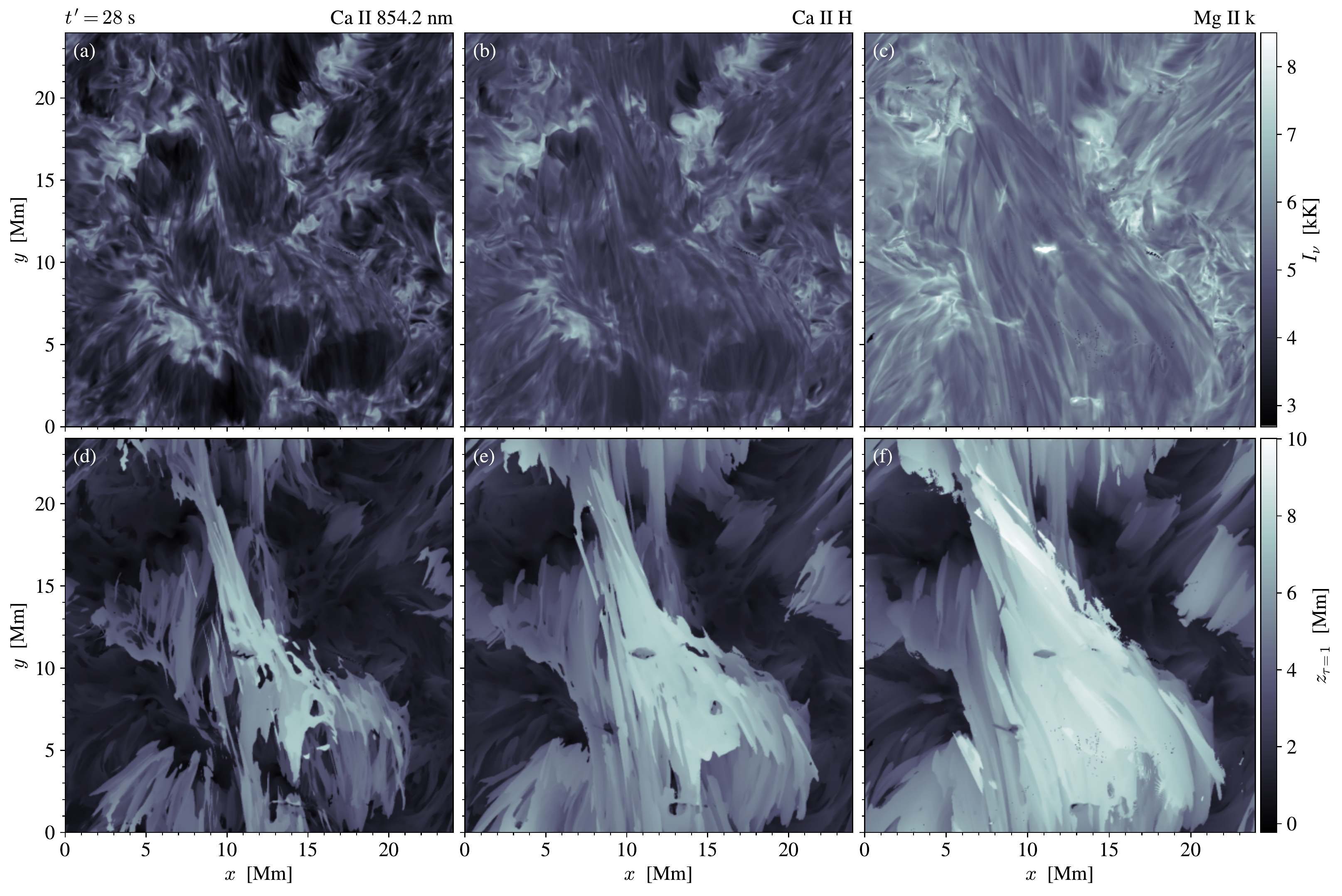}
    \caption{\ion{Ca}{ii}~854.2~nm, \ion{Ca}{ii}~H, and \ion{Mg}{ii}~k nominal line core intensity (upper panels) and $\tau = 1$ heights (lower panels) at $t^\prime = 28$~s. The colourbar of the intensity is clipped at 8.5~kK to emphasise the less bright features of the \ion{Ca}{ii} line cores.}
    \label{fig:lam_tau}
\end{figure*}

Figure~\ref{fig:multi_int} shows that the strongest emission of the different spectra is found at L2. This location is promising in terms of electron beam heating, as it is located at the footpoint that is connected to the longest coronal loops. The upper panel in Fig.~\ref{fig:sel_areas} shows a large number of electron acceleration sites that are connected to the particular magnetic field footpoint. Even though the atmospheric response to the electrons has proven difficult to single out, the TR and chromospheric spectra might still be affected by the electrons depositing their energy along the loops. Figure~\ref{fig:atmos_resp} (n) shows that the electrons along the line of sight deposit most of their energy at TR and chromospheric heights, hence it is possible that the increased intensity seen in the L2 column of Fig.~\ref{fig:multi_int} is caused by local heating events. 

\subsection{Line formation}

Figure~\ref{fig:lam_tau} shows the \ion{Ca}{ii}~854.2~nm, \ion{Ca}{ii}~H, and \ion{Mg}{ii}~k line cores (top row) and $\tau = 1$ heights (bottom row) for the entire simulation domain at a single snapshot in time ($t^\prime = 28$~s). For simplicity, the line core is defined at $v_\mathrm{D} = 0$ (see the dashed line in Fig.~\ref{fig:multi_int}), even though the concept of a single line core vanishes when analysing complex atmospheres with multi-layered structures. We have chosen one of the later Bifrost snapshots so that the electrons will have affected the atmosphere for a longer duration (the electrons are present from $t^\prime = 0$~s). Panels (a)--(c) show emission of the line cores at the magnetic field footpoints. The prominent current sheet at $(x, y) = (10, 11)$ has enhanced intensity in the line cores, see \cite{2019A&A...626A..33H} 
for a detailed analysis of the associated Ellerman bomb and UV burst emission. The emission from the spectral lines consists of long strands outlining the loops above the flux emergence region. This region is seen in panels (d)--(f) as the structure with the highest $z_{\tau = 1}$ values. The structure is not as clearly outlined in panel (d) as in the other panels. This is because \ion{Ca}{ii}~854.2~nm forms deeper in the atmosphere, hence there is less emission of the line core above the flux emergence region. This is also seen in panel (a), where there are fewer long emission strands outlining the magnetic bubble compared to the other upper panels. The \ion{Ca}{ii}~H and \ion{Mg}{ii}~k line cores form higher in the chromosphere, hence larger portions of the flux emergence region are outlined in the lower panels. 

\begin{figure*}[!ht]
    \centering
    \includegraphics[width=\textwidth]{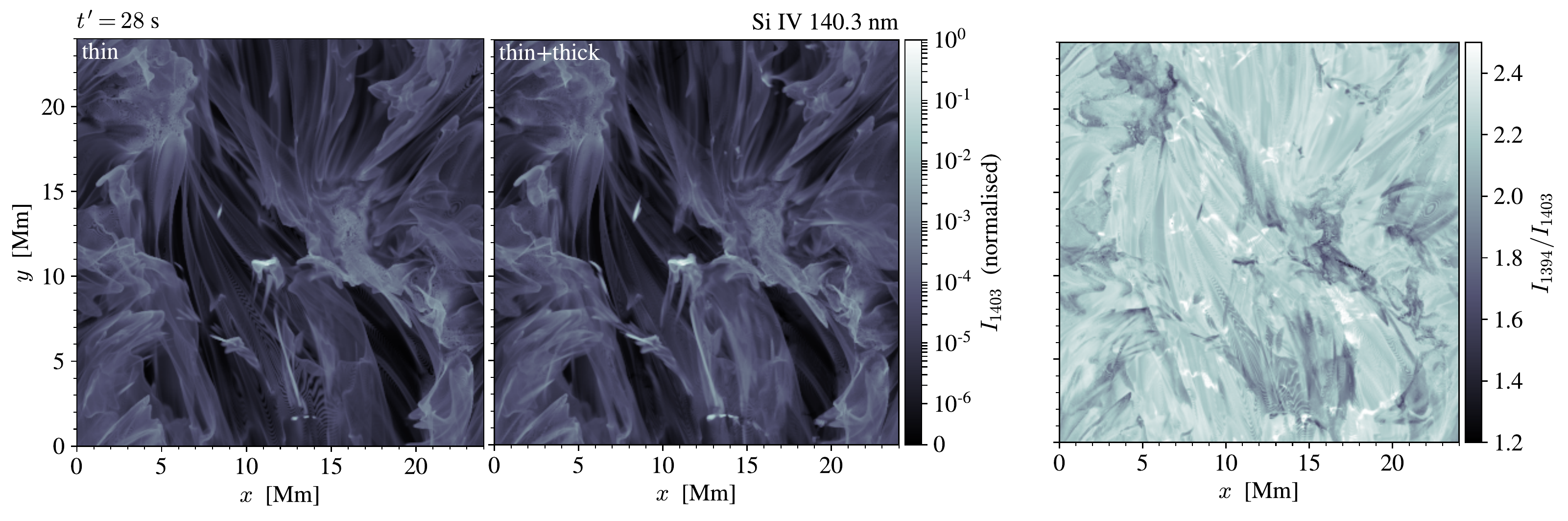}
    \caption{Integrated \ion{Si}{iv}~140.3~nm intensity calculated using CHIANTI (left, marked as optically thin) and RH1.5D (middle, marked as optically thin+thick), and ratio of the \ion{Si}{iv} resonance lines (right). The three panels show their respective quantities at $t^\prime = 28$~s. The integrated intensity maps are normalised between 0 and 1. The $I_{1394}/I_{1403}$ ratio is calculated using the intensity output from the RH code.}
    \label{fig:si_compare}
\end{figure*}

Figure~\ref{fig:si_compare} shows the integrated \ion{Si}{iv}~140.3~nm intensity calculated using two different line synthesis approaches, as well as the ratio of the \ion{Si}{iv}~139.4~nm to \ion{Si}{iv}~140.3~nm line. In the left panel, the emission is calculated employing an optically thin approach using CHIANTI. The middle panel shows the intensity as output from the RH1.5D code allowing the \ion{Si}{iv} resonance lines to form under both optically thin and thick conditions.  
The intensity in both panels is normalised between 0 and 1 as we aim to do a qualitative comparison between the two intensity maps. A detailed quantitative comparison is difficult given differences between the two approaches in for example silicon abundance and temperature coverage. Hence Fig.~\ref{fig:si_compare} aims to visualise the impact of using a model atom that includes potential opacity effects. At first glance, the integrated intensity maps look similar. The general structure of the simulation is outlined by the intensity in both panels, with long strands spanning the flux emergence region. However, the structures are smoother in the left panel where the emission forms under optically thin conditions. The middle panel has features that appear to be below the loops that outline the flux emergence region. These features are either weak or not seen in the left panel, suggesting that it is necessary to include potential opacity effects when calculating \ion{Si}{iv} synthetic spectra. 

This is further supported by the right panel, which shows the ratio of the \ion{Si}{iv}~139.4~nm to \ion{Si}{iv}~140.3~nm line. In the optically thin limit the ratio should be equal to two, which is the ratio of their oscillator strengths. While the figure shows that most of the \ion{Si}{iv} lines form under optically thin conditions, there are also darker areas where the ratio is below two. Similar results were discovered in \cite{2023A&A...672A..47S}, 
where the wavelength integrated \ion{Si}{IV} ratio was found to be between 1.6 and 1.8 at four different locations in a loop-like structure in a MURaM simulation. Our results show that a few of the darker areas where the ratio is below two stretches along the strands above the flux emergence region, which is consistent with their findings. Our results underline the risk of assuming that all \ion{Si}{iv} emission forms under optically thin conditions in the solar atmosphere, and motivates our choice of using a more advanced approach when calculating the \ion{Si}{iv} synthetic spectra.

\subsection{Contribution function to the intensity}

\begin{figure*}[ht]
    \centering
    \subfigure{\includegraphics[width=\columnwidth]{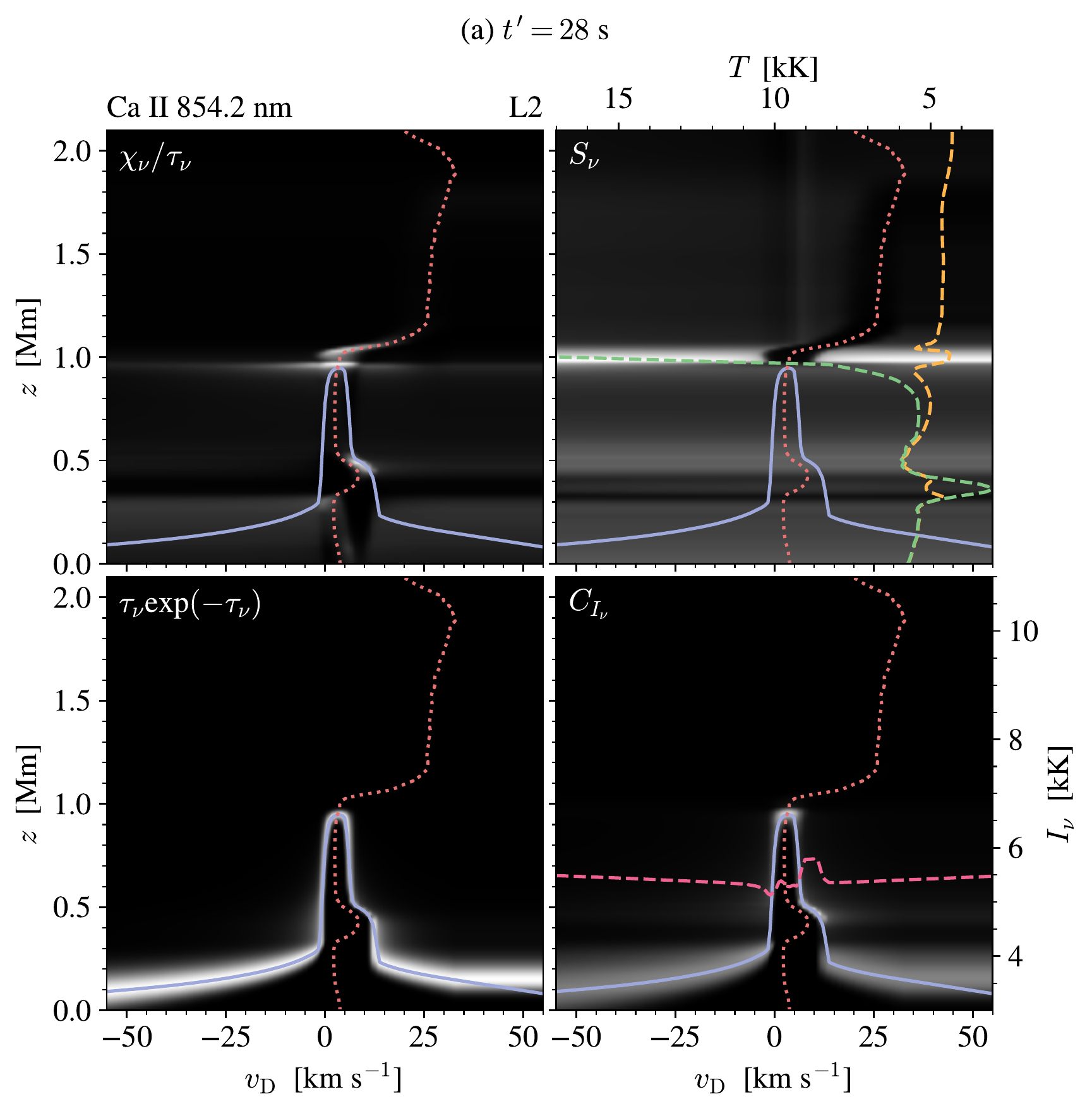}}
    \subfigure{\includegraphics[width=\columnwidth]{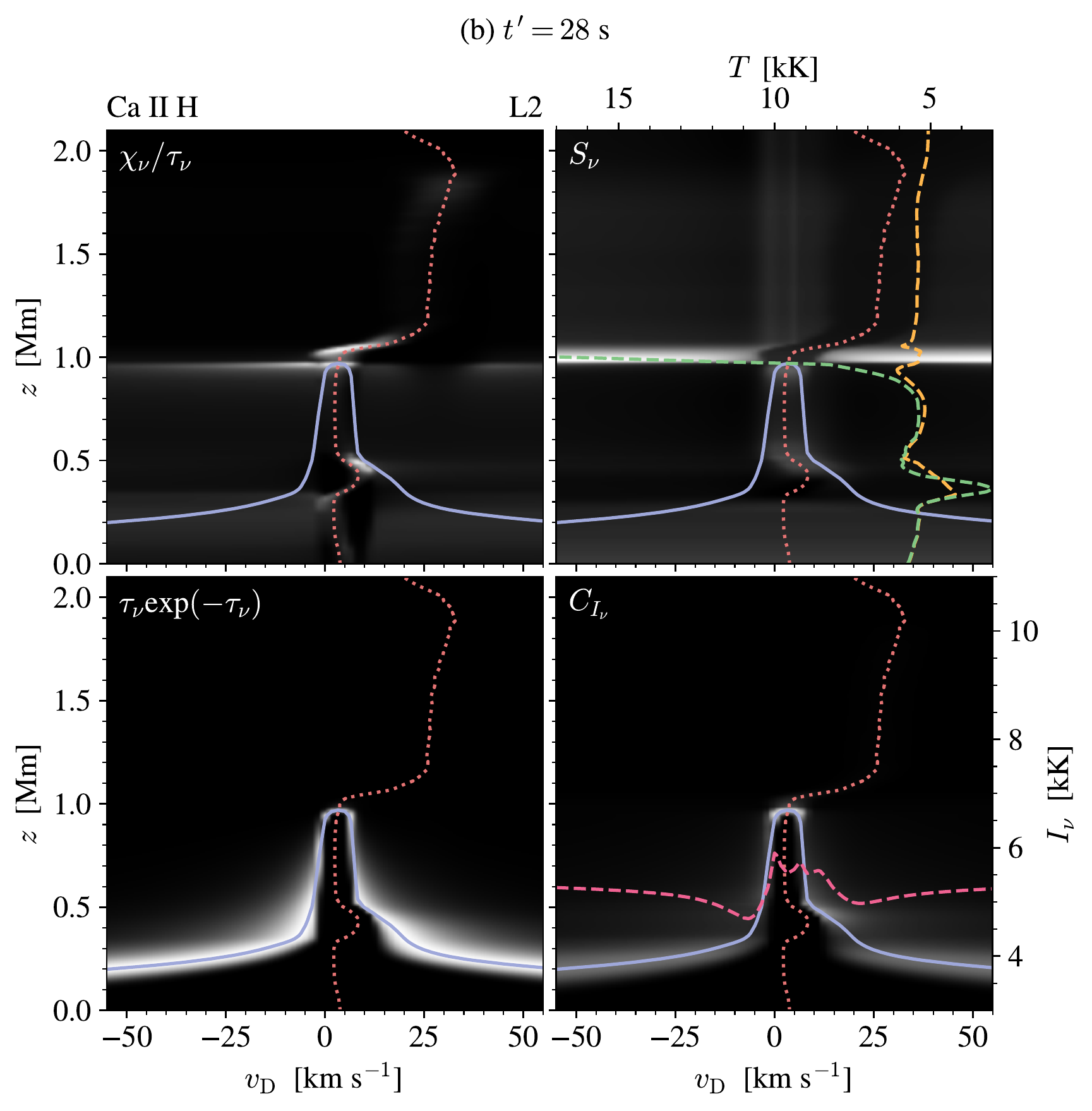}}
    \subfigure{\includegraphics[width=\columnwidth]{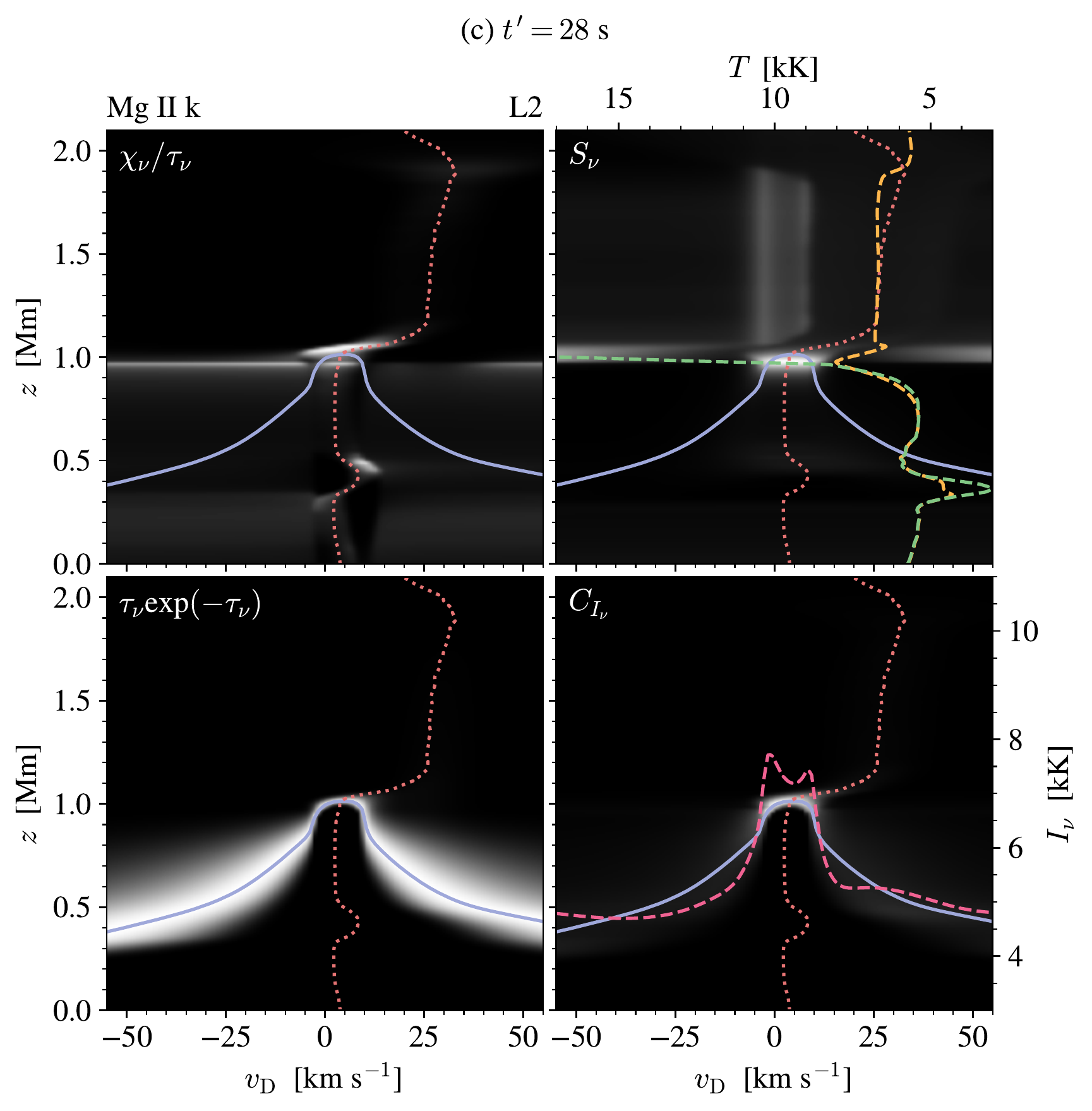}}
    \subfigure{\includegraphics[width=\columnwidth]{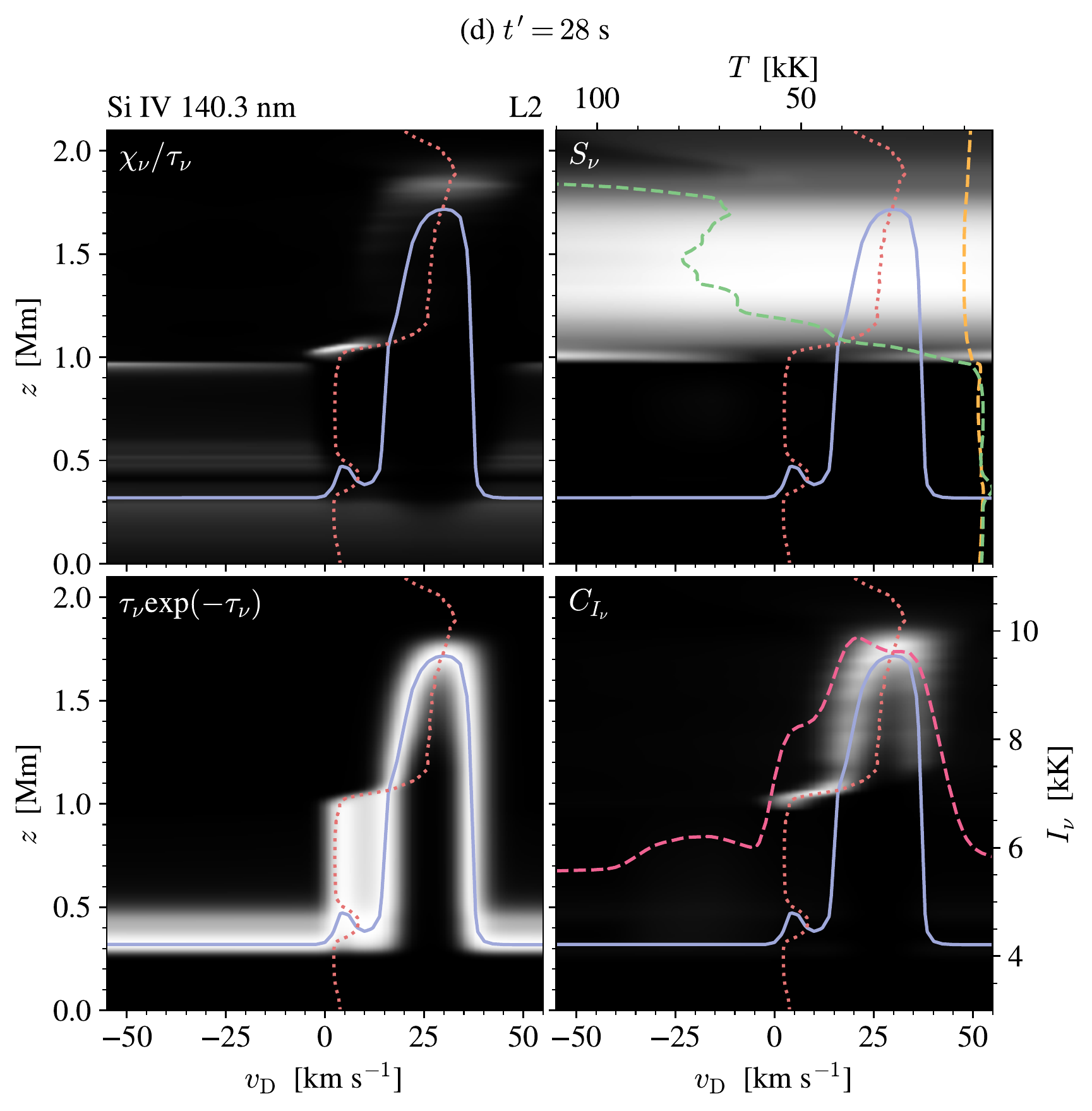}}
    \caption{Intensity formation of the \ion{Ca}{ii}~854.2~nm (a), \ion{Ca}{ii}~H (b), \ion{Mg}{ii}~k (c), and \ion{Si}{iv}~140.3~nm (d) spectral lines at the L2 location at $t^\prime = 28$~s. Each subfigure consists of four panels, where the quantities given in the top left corners are shown in greyscale as functions of frequency from the line centre (in units of Doppler offset) and atmospheric height. The $\tau_\nu = 1$ height (purple solid) and vertical velocity (red dotted) are displayed in all panels. Negative (positive) velocities correspond to upflows (downflows). The top right panels display the source function at $v_\mathrm{D} = 0$ (yellow dashed) and Planck function (green dashed) in units of brightness temperature specified along the top (we note that the temperature range in (d) is larger because \ion{Si}{iv} is sensitive to much higher temperatures). Multiplication of the first three panels produces the contribution function in the bottom right panel. This panel also contains the intensity profile (pink dashed) in units of brightness temperature. Gamma correction is added to the $C_{I_\nu}$ term to amplify the weaker values.}
    \label{fig:4pL2} 
\end{figure*}

\begin{figure*}[ht]
    \centering
    \subfigure{\includegraphics[width=\columnwidth]{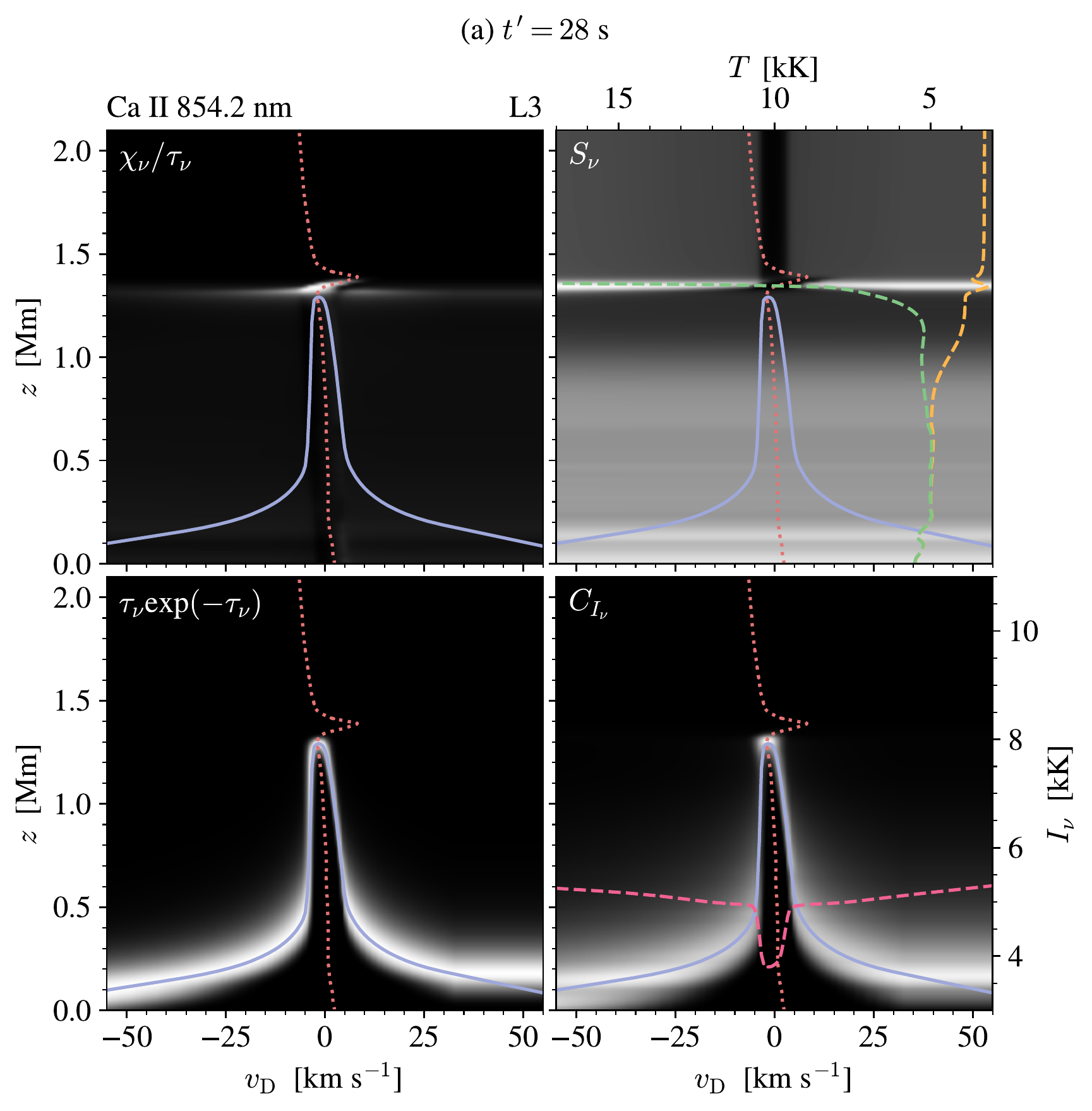}}
    \subfigure{\includegraphics[width=\columnwidth]{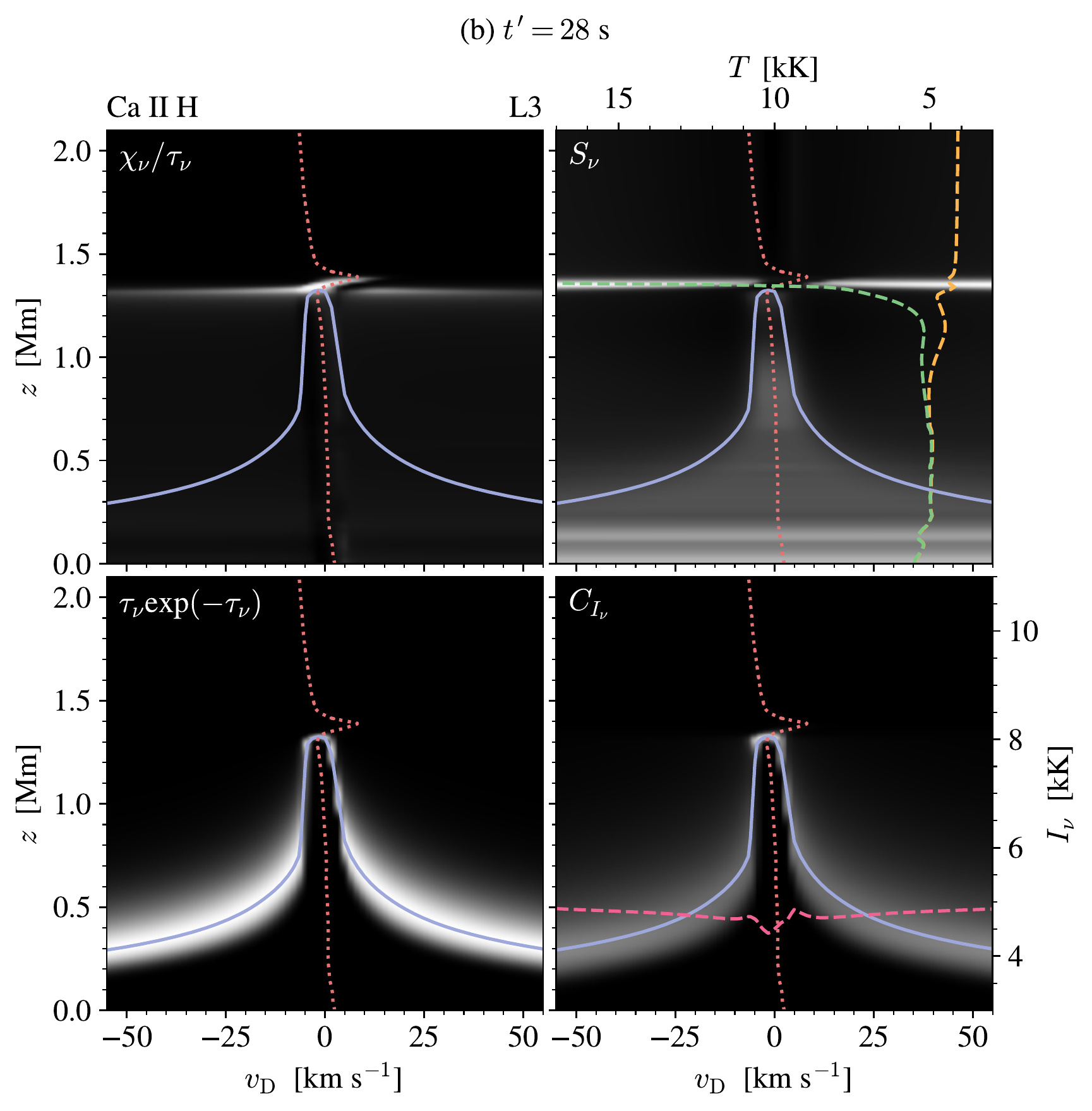}}
    \subfigure{\includegraphics[width=\columnwidth]{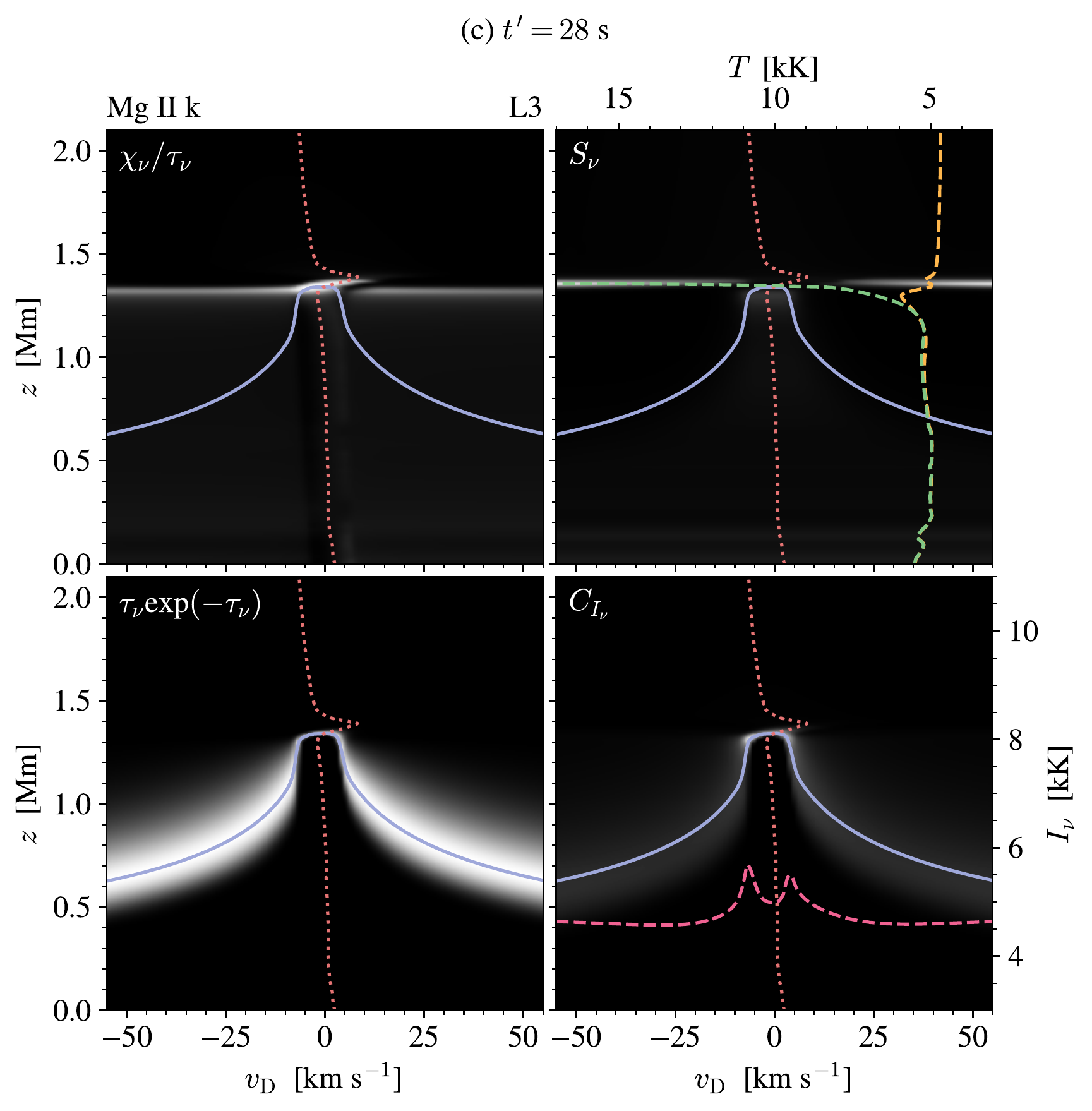}}
    \subfigure{\includegraphics[width=\columnwidth]{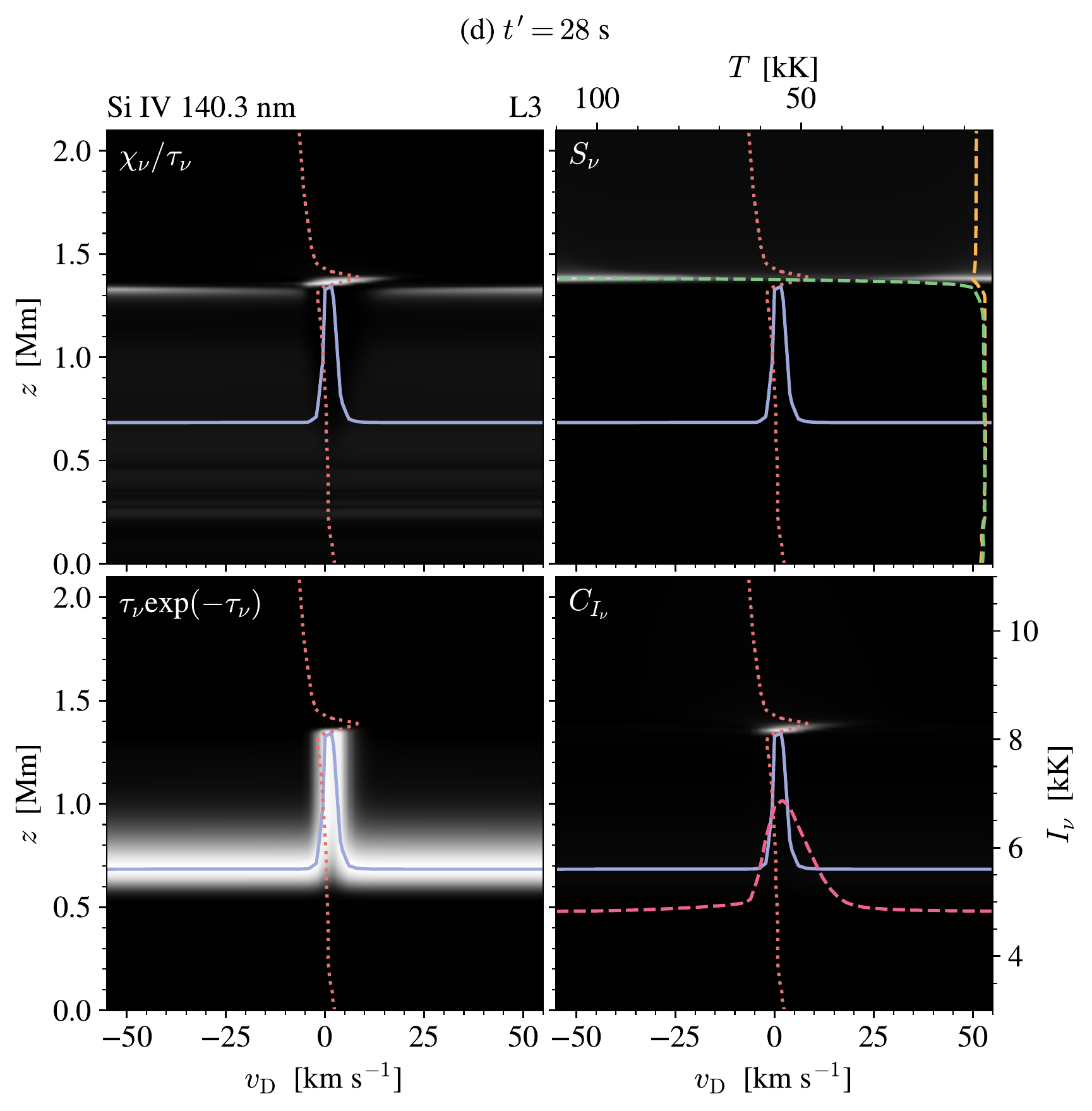}}
    \caption{Intensity formation of the \ion{Ca}{ii}~854.2~nm (a), \ion{Ca}{ii}~H (b), \ion{Mg}{ii}~k (c), and \ion{Si}{iv}~140.3~nm (d) spectral lines at the L3 location at $t^\prime = 28$~s. See the caption of Fig.~\ref{fig:4pL2} for more details.} 
    \label{fig:4pL3} 
\end{figure*}

\begin{figure*}[ht]
    \centering
    \subfigure{\includegraphics[width=\columnwidth]{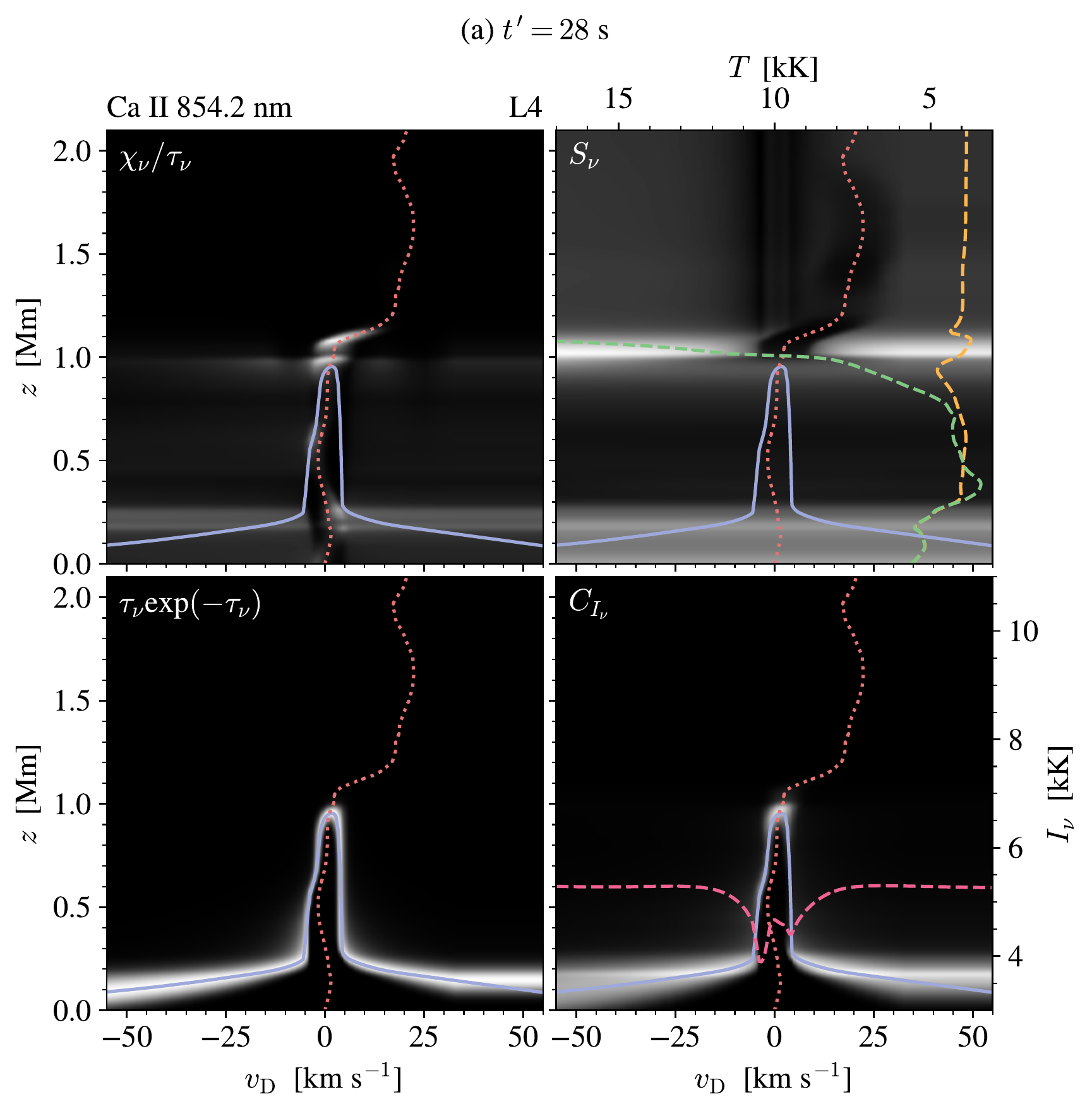}}
    \subfigure{\includegraphics[width=\columnwidth]{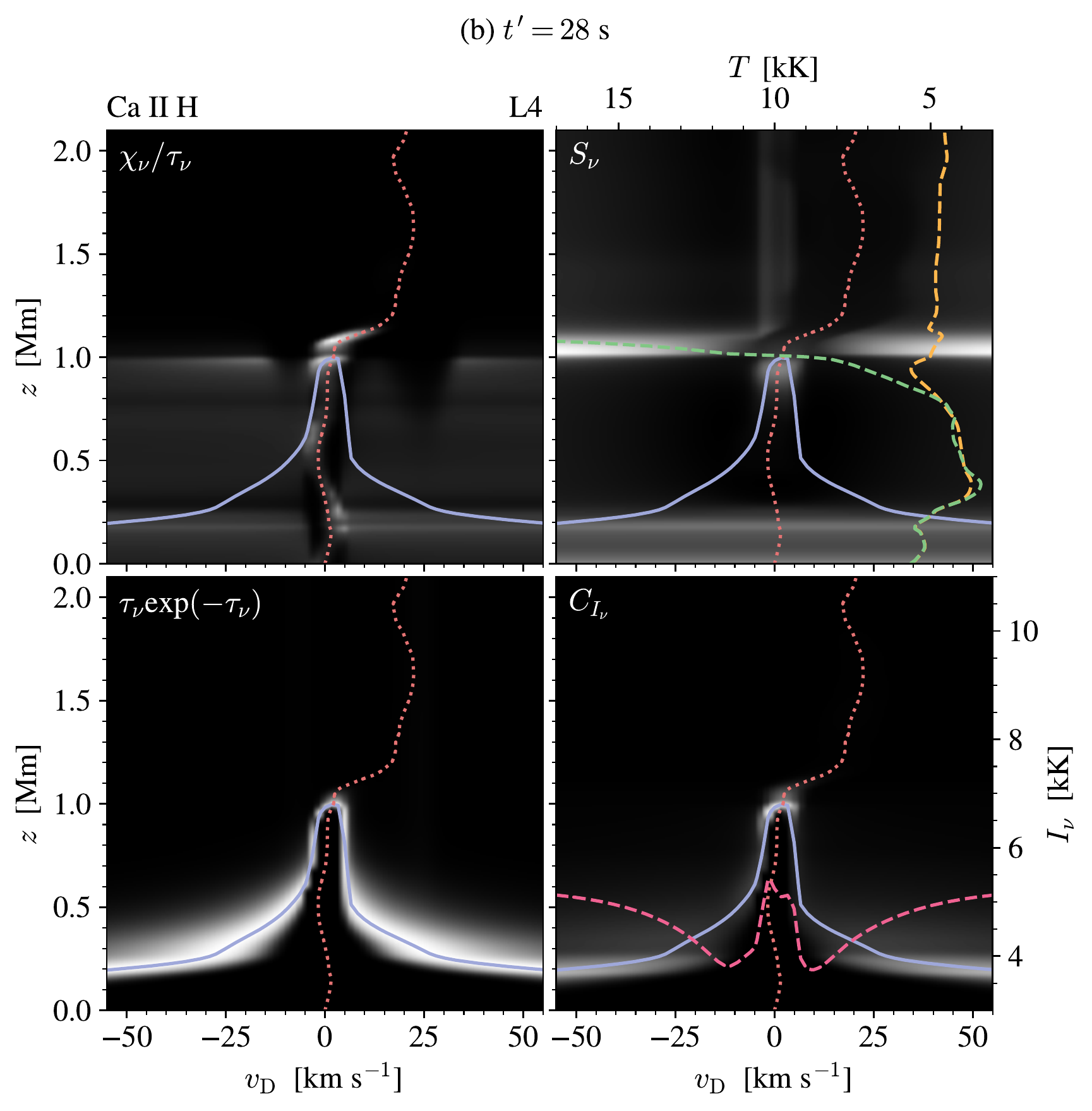}}
    \subfigure{\includegraphics[width=\columnwidth]{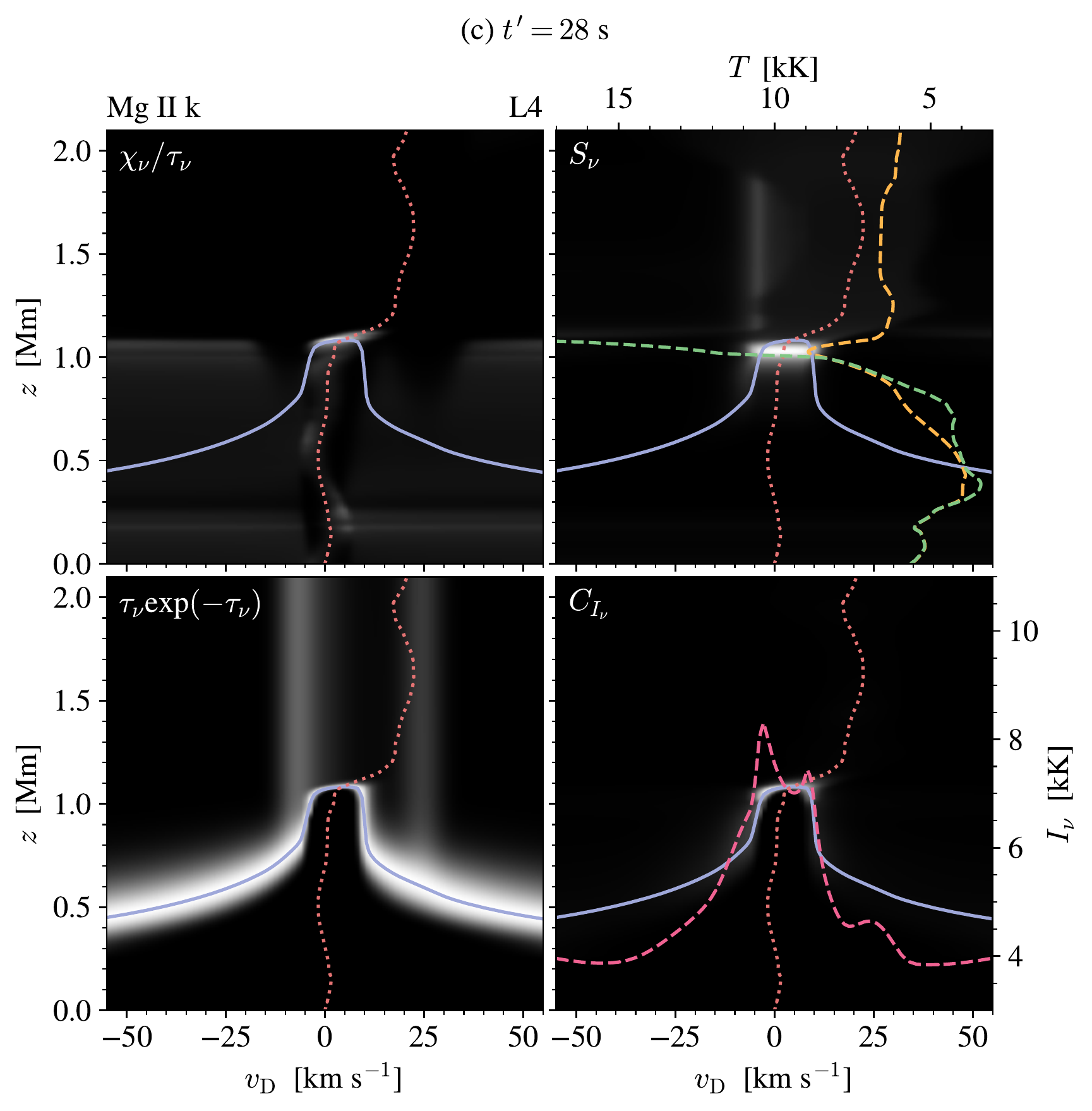}}
    \subfigure{\includegraphics[width=\columnwidth]{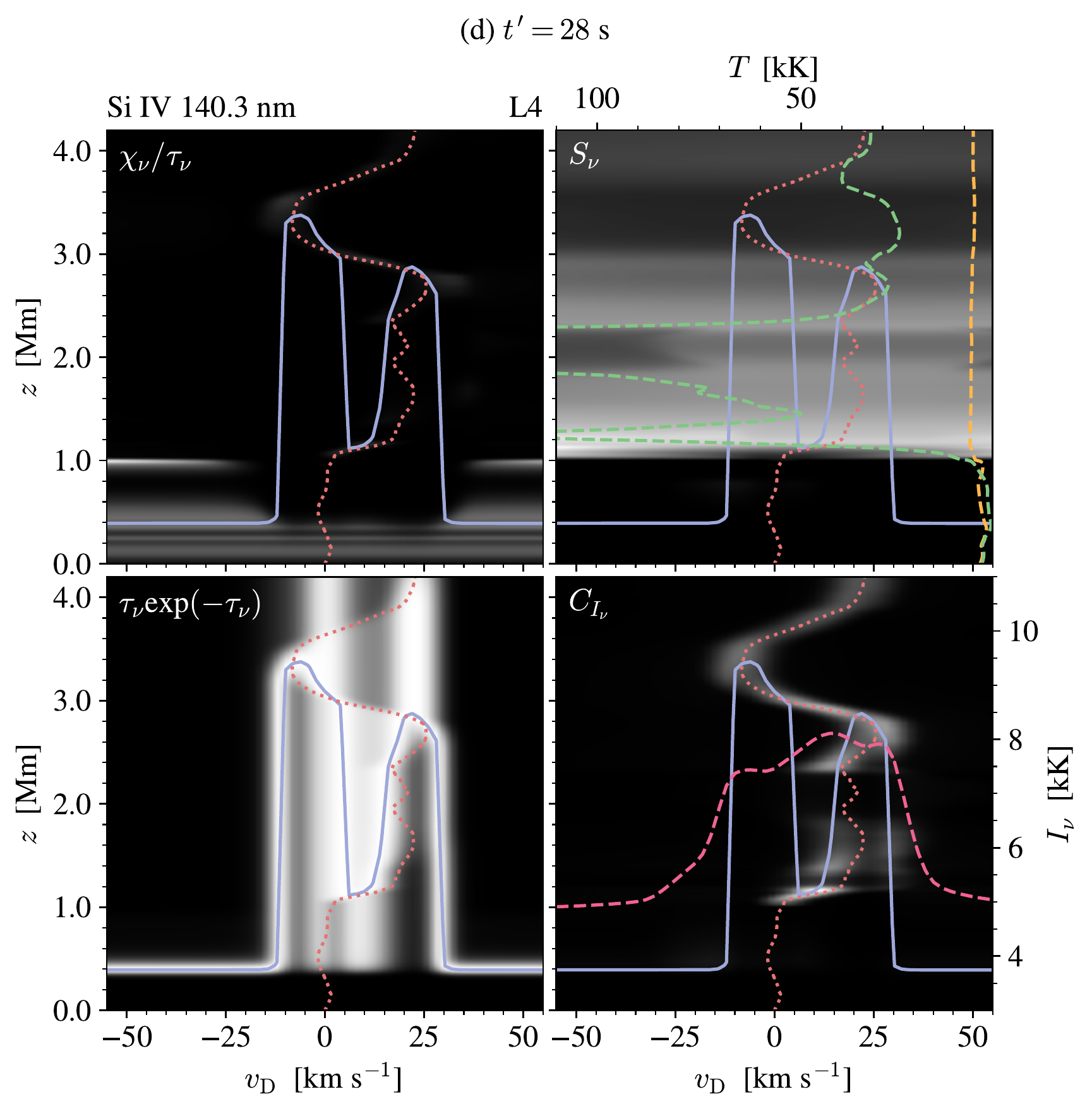}}
    \caption{Intensity formation of the \ion{Ca}{ii}~854.2~nm (a), \ion{Ca}{ii}~H (b), \ion{Mg}{ii}~k (c), and \ion{Si}{iv}~140.3~nm (d) spectral lines at the L4 location at $t^\prime = 28$~s. We note that the maximum height in (d) is larger than the other panels. See the caption of Fig.~\ref{fig:4pL2} for more details.} 
    \label{fig:4pL4} 
\end{figure*}

Figures~\ref{fig:4pL2}--\ref{fig:4pL4} show four $2 \times 2$ diagrams of the intensity formation of \ion{Ca}{ii}~854.2~nm, \ion{Ca}{ii}~H, \ion{Mg}{ii}~k, and \ion{Si}{iv}~140.3~nm at the L2, L3, and L4 locations at $t^\prime = 28$~s. The panels in each subfigure represent the individual terms in Eq.~\ref{eq:cont_func} as well as the total contribution function to the line intensity.

Panels (a) and (b) in Fig.~\ref{fig:4pL2} show the intensity formation of \ion{Ca}{ii}~854.2~nm and \ion{Ca}{ii}~H, respectively. There are two distinct downward velocity gradients that are reflected in the $\chi_\nu/\tau_\nu$ term around 0.5~Mm and 1.05~Mm. The velocity gradient at 1.05~Mm seems to be located just above the maximum formation height of both spectral lines, and does not affect the contribution to the line intensity. The velocity gradient at 0.5~Mm (see also Fig.~\ref{fig:atmos_resp} (b)) causes small emission features in the red wing, which can be seen in the emergent intensity profile that is shown in the lower right panels. The source function is coupled to the Planck function from the photosphere up to 0.3~Mm, where a narrow cold region is seen as a dip in the Planck function and as a darker band in $S_\nu$. Above the cold region, the source function is more closely following the Planck function again. For \ion{Ca}{ii}~854.2~nm, the functions decouple around 0.6~Mm, while for \ion{Ca}{ii}~H this does not happen until 0.8~Mm.  
There is a strong increase in the Planck function, and hence also temperature, around 1~Mm. This gives rise to an increase in the source function around the maximum height of formation for the two lines. In panel (a), the increase in source function is responsible for the central reversal of the \ion{Ca}{ii}~854.2~nm line core ($v_\mathrm{D} = 0$), while the absorption feature at $-1$~km~s$^{-1}$ is caused by the decline in source function just above the narrow cold band. In panel (b), the increase in source function gives rise to two emission peaks in the \ion{Ca}{ii}~H intensity profile. The line core is formed at the maximum height of the optical depth unity curve ($z = 1$~Mm), where the central reversal is caused by the source function declining from the peak at around 0.95~Mm.

Figure~\ref{fig:4pL2} (c) shows the intensity formation of \ion{Mg}{ii}~k. The velocity gradients at 0.5~Mm and 1.05~Mm are reflected in the $\chi_\nu/\tau_\nu$ term. The gradient at 0.5~Mm occurs below the formation height of the spectral profile, while the gradient at 1.05~Mm does not affect the total contribution function significantly because the other terms are too small. In turn, the contribution function is dominated by the peak in source function, which is caused by the sudden increase in temperature seen as the Planck function. The central emission is caused by the increasing source function, while there is a shallow central absorption caused by a declining source function in line centre. 

Figure~\ref{fig:4pL2} (d) shows the formation of the \ion{Si}{iv}~140.3~nm line. The downward velocity gradient at 1.05~Mm is clearly seen in the $\chi_\nu/\tau_\nu$ term. The $\tau = 1$ height reaches high altitude, up to 1.7~Mm, at about $+35$~km~s$^{-1}$ Doppler offset. We note that all the spectral profiles presented in Fig.~\ref{fig:4pL2} are shifted to the red to varying degree. The strongest redshift is seen in \ion{Si}{iv}, which has an average redshift of $+6$~km~s$^{-1}$ in the entire box. This is due to the overall positive velocity at the formation height shifting the profile to the red. In panel (d), the strong velocity gradient also causes a broadening of the asymmetric profile. The contribution function panel in the lower right shows that the part of the profile formed above approximately 1~Mm is formed under optically thick conditions, as this is where the $\tau = 1$ curve follows the peak of the contribution function. Below 1~Mm on the other hand, the $\tau = 1$ curve departs from the contribution function, which means that the formation for this part of the line is under optically thin conditions.

Figure~\ref{fig:4pL3} represents the intensity formation of the four different spectral lines at the L3 location. Panel (a) shows that \ion{Ca}{ii}~854.2~nm is less affected by PRD than the other lines, as the source function, which appears like a horizontal band, shows little variation along the $x$-axis. The maximum height of the optical depth unity curve is slightly below the height of both the downward velocity gradient and the sudden temperature increase, hence these features do not contribute to the intensity and are not reflected in the emergent intensity profile. The intensity of the line is therefore just a map of the source function at optical depth unity. The source function decreases with height up to where the line core forms, causing the overall absorption profile without emission features.

Panels (b) and (c) in Fig.~\ref{fig:4pL3} show the formation of \ion{Ca}{ii}~H and \ion{Mg}{ii}~k, respectively. The maximum height of the \ion{Ca}{ii}~H optical depth unity curve is just below that of \ion{Mg}{ii}~k. The \ion{Ca}{ii}~H profile has two peaks as a result of the peak in source function. There is an increase in the source function just after it decouples from the Planck function at approximately 0.7~Mm that gives rise to a small intensity increase in the red peak. The velocity gradient also makes a small contribution to the emission of the red peak, which has a slightly higher intensity than the blue peak. The rest of the line profile forms similarly as \ion{Ca}{ii}~854.2~nm, where the intensity maps the source function. The red and blue peaks of the \ion{Mg}{ii}~k profile are caused by the velocity gradient and temperature increase around 1.35~Mm, while the absorption feature is caused by the decline in the source function at the maximum height of the optical depth unity curve. The source function decouples from the Planck function at a higher height (around 1.25~Mm) compared to \ion{Ca}{ii}~H, giving a larger rise in source function that leads to more pronounced peaks.

Figure~\ref{fig:4pL3} (d) shows the intensity formation of \ion{Si}{iv}~140.3~nm. The contribution function is dominated by the $\chi_\nu/\tau_\nu$ term, which is caused by the weak downflowing velocity around $z = 1.35$~Mm. The velocity causes the single peaked profile to shift to the red. We know from Fig.~\ref{fig:multi_int} that the lines forming at the L3 location are subject to shock waves passing through the atmosphere, where the oscillations in temperature contributes to increases and decreases in intensity. The $2 \times 2$ diagrams only show a single instance in time, and at $t^\prime = 28$~s the maximum formation height of the \ion{Si}{iv} line is just below the height of the temperature increase. Hence we can assume that the line is formed between the shock waves.

Figure~\ref{fig:4pL4} shows the formation of the different spectral profiles at the L4 location. The panels in (a) represent the intensity formation of \ion{Ca}{ii}~854.2~nm. The line forms below the height of both the velocity gradient at 1.1~Mm and the temperature increase at 1~Mm. There is a decrease in the source function after it decouples from the Planck function. When the temperature starts to increase around 0.7 Mm, the source function increases too. This, together with the increase in $\chi_\nu/\tau_\nu$ at the maximum $\tau = 1$ height, causes an emission feature in the line core. 

Panels (b) and (c) in Fig.~\ref{fig:4pL4} show the intensity formation of \ion{Ca}{ii}~H and \ion{Mg}{ii}~k, respectively. Both lines are double peaked, with a blue peak that has higher intensity than the red peak. The two peaks in the \ion{Ca}{ii}~H profile are caused by the increase in source function around 0.95~Mm. The blue peak is more intense because the $\chi_\nu/\tau_\nu$ term gives a stronger contribution on the blue side. The decline in source function at the highest height of the $\tau = 1$ curve causes the absorption feature at approximately $+2$~km~s$^{-1}$. We note that there is a strong velocity gradient around 1.1~Mm that does not contribute to the formation of the \ion{Ca}{ii}~H line intensity. However, the \ion{Mg}{ii}~k line forms at this height, and the strong downward velocity gradient causes the line core to shift to the red. The blue peak is more intense than the red peak because it gets a larger contribution from the $\tau_\nu \mathrm{exp}(-\tau_\nu)$ term, but both peaks get a significant contribution from the increase in temperature and source function at 1.05~Mm. The absorption feature is caused by the decline in source function at 1.1~Mm. The increased emission of the red wing (around $+25$~km~s$^{-1}$) and blue wing (around $-10$~km~s$^{-1}$) is most likely caused by the bright columns seen in the $\tau_\nu \mathrm{exp}(-\tau_\nu)$ panel, even though the contribution is too small to be visible in the $C_{I_\nu}$ panel.

Figure~\ref{fig:4pL4} (d) shows the formation of the \ion{Si}{iv}~140.3~nm line. The complex structure of the atmosphere at L4 (see Fig.~\ref{fig:atmos_resp} (d)) results in the line features forming at very different heights (we note that the height in Fig.~\ref{fig:4pL4} (d) ranges from 0 to 4~Mm, whereas the height in the other panels ranges from 0 to 2~Mm). We can distinguish both a blue and a red component in the line profile. These are formed at $z = 3$ and $z = 3.2$~Mm, where the $\tau = 1$ heights have peaks at $-10$ and $+25$~km~s$^{-1}$. These velocity components contribute to the broadening of the spectral profile. The line core forms around 1.05~Mm, where there is a downward velocity component that causes it to shift to the red. The emission of the line core is caused by the sudden increase in temperature and source function. 

\section{Discussion}

The analysis presented in this work shows that the chromospheric and TR spectra are highly affected by strong velocity gradients and sudden variations in temperature. It is difficult to determine if these variations are due to the non-thermal electrons depositing their energy along the magnetic field, especially since the simulation is multi-dimensional and potential effects that occur are not aligned with the particular vertical columns used for calculating the emergent spectra. Even though we cannot make a firm conclusion about the effect the non-thermal electrons have on the synthetic spectra, our results contribute to the continued pursuit of understanding small-scale reconnection events and their impact on the solar atmosphere.

To determine signatures in the synthetic spectra that may arise from the non-thermal electrons, we studied the evolution of the atmosphere and the response to the accelerated electrons. \cite{2020A&A...643A..27F} 
have shown that the energy transport by accelerated electrons and thermal conduction differs greatly with depth in the lower atmosphere. Heating by thermal conduction dominates at TR heights, but decreases towards the chromosphere due to the temperature drop. The non-thermal electrons are not directly affected by the sudden decrease in temperature at TR heights, and beam heating generally exceeds conductive heating in the chromosphere. This means that synthetic spectra forming at TR heights, such as the \ion{Si}{iv} resonance lines, are likely to be affected by both electron beam heating and thermal conduction, while the synthetic chromospheric spectra should be mostly affected by the non-thermal electrons. However, Fig.~\ref{fig:atmos_resp} shows that there is no clear indication that the electron beams are affecting the evolution of the atmosphere. This is most likely due to the low value of $E_\mathrm{C}$ for the non-thermal electrons in the corona, but also because the energy transport is very low. Small values of $E_\mathrm{C}$ (around 1--2~keV) implies that the effect of non-thermal electrons on the TR and chromosphere are similar to that of thermal conduction \citep{2014Sci...346B.315T, 2018ApJ...856..178P, 2020ApJ...889..124T}. 
Additionally, signatures from non-thermal electrons in chromospheric spectra greatly diminishes when the electrons deposit their energy in the corona.
In an attempt to add maximum power to the electron beams, we performed an experiment where all the energy from the reconnection events was transferred to the electrons ($p = 1$). In this experiment, the atmospheric structure was almost identical to the original simulation where $p = 0.2$, and the impact on the spectral diagnostics was insignificant. The only notable difference was in the beam heating, which was increased by a factor of $5$. This tells us that the beam heating events in this simulation are too weak to significantly affect the low atmosphere, even when the electron beams carry the maximum amount of energy that is possible in this simulation. 

The low level of change in our simulation might be due to the relatively short time that the electrons are present. 
\cite{2022A&A...668A.177R} 
have demonstrated that it takes approximately 800~s (from the magnetic field is ordered) for the field in a Bifrost simulation of the quiet Sun to generate enough magnetic energy to produce heating events of typical nanoflare energies ($10^{24}$~erg). \cite{2017A&A...603A.103G} 
have shown that most reconnection events in a Bifrost simulation similar to ours have lifetimes of roughly 40~s, with a weighted average of around 50--60~s. During that lifetime, the energy released by the small-scale events typically ranges from $10^{20}$--$10^{24}$~erg, which is the same as what \cite{2020A&A...643A..27F} 
have predicted for longer lasting electron beam heating events. Our 36~s of simulation time is of the order of the shortest events presented in similar simulations, as the high computational cost has so far limited the running time. A longer simulation, including more full time-scale heating events would most likely produce more locations in the chromosphere where the effect of the electron beams would leave their imprint. At this point there is no plan to shoulder the computational cost required without also changing the solar environment to a more active region. To say with certainty that a strong signal would show up in this simulation by running it for longer time cannot therefore be guaranteed. 

The travel distance from the site of reconnection to the site of the deposited electron energy is affected by the power-law index $\delta$. A low power-law index allows the electrons to penetrate deeper into the atmosphere, while larger values lead to energy being deposited higher in the atmosphere. This is because the rate of deposited energy increases more rapidly for larger values of $\delta$, meaning that the amount of energy deposited in the lower atmosphere is less than for smaller values of $\delta$. Consequently, the spectra forming higher in the atmosphere, such as the \ion{Si}{iv} resonance lines, are more likely to be affected by the non-thermal electrons compared to spectra forming at lower heights. Generally, we expect to see a large difference in the intensity and shape of the spectral lines when the time offset between the non-thermal electrons and the thermal conduction front is the greatest. In reality, this happens if a reconnection event occur high in the atmosphere, meaning that a relatively large amount of energy is transferred to the electrons and the electrons travel a great distance. This comes from the fact that travel distance increases linearly with height, while the available energy decreases exponentially with height. In this analysis, we have chosen columns that are situated at the magnetic field footpoints of the simulation. Even though L2, L3, and L4 are connected to field lines showing large changes in average electron beam heating power, we do not know if energetic events in the corona have an effect on the lower atmosphere. The most significant beam heating is seen at L2 (Fig.~\ref{fig:atmos_resp} (n)), where energy from the non-thermal electrons is deposited at TR and chromospheric heights. What is unique about L2 is that the electrons responsible for the peak in beam heating around 1~Mm are not accelerated from local reconnection in the lower atmosphere, as we do not see negative values of $Q_\mathrm{b}$ of the same magnitude at approximately the same height. At L3 and L4, the electrons deposit their energy almost immediately after they are accelerated. At L2, there are two acceleration sites (at $z = 2$ and 9~Mm) where energy is transferred to the electrons. However, since the angle between the magnetic field and the vertical direction at these heights differs from the angle at 1~Mm, these events are not related. It is therefore possible that the electrons depositing their energy at this height might be accelerated from reconnection events in the corona.

L2 is the most promising location in terms of signatures from non-thermal electrons. The electron energy deposited at 1~Mm is consistent with the upflows of hot plasma into the TR and corona, and further takes place around the formation height of the \ion{Ca}{ii} and \ion{Mg}{ii} lines. However, it is difficult to know if the strong velocity gradient is caused by the electrons depositing their energy at this height, especially since we see velocity gradients that are consistent with the sudden temperature increase from the chromosphere to the TR at all four locations. Additionally, the velocity gradients at L3 and L4 do not seem to be directly affected by the deposited electron energy, which gives reason to believe that this is not the case at L2 either. This is further supported by comparing the spectral lines, where the shape and features of the \ion{Mg}{ii}~k line at L2 and L4 are both caused by steep velocity gradients and sudden increases in temperature. If the electrons have a significant impact on the temperature and velocity at L2, we expect to see a larger difference between the spectral profiles at the two locations. However, we cannot be certain that the electrons do not affect the atmosphere, and hence also the spectra, even though there are no significant effects from the energy deposited directly in the TR and chromosphere. We know that the electrons are continuously accelerated throughout the simulation, and they might be affecting the result more passively compared to larger energy releases. The \ion{Ca}{ii}~H, \ion{Mg}{ii}~k, and \ion{Si}{iv}~104.3~nm lines show similarities to those produced by some of the RADYN models in \citet{2018ApJ...856..178P}, 
\citet{2020ApJ...889..124T}, 
and \citet{2022A&A...659A.186B}, 
in particular the low-temperature (1~MK) loop models. The similarities include increased emission of the \ion{Ca}{ii}~H and \ion{Mg}{ii}~k blue peaks, slight redshift of the \ion{Mg}{ii}~k line core, emission of the blue wing of \ion{Mg}{ii}~k, and single peaked \ion{Si}{iv}~104.3~nm profiles that are strongly redshifted. The fact that we see spectral features that are similar to the signatures caused by non-thermal electrons in RADYN models suggests that the accelerated electrons in the Bifrost simulation have an impact on the atmosphere. However, even though we have an idea of the mechanisms behind small-scale heating events and the transport of non-thermal electrons, it is difficult to make a conclusion from our simulation without observational proof. 

The results of the spectral line analysis can give an indication of what to look for in observations. The non-thermal electrons present in the Bifrost simulation might have an impact on the atmosphere, even though spectral line features that arise as a consequence to the beam heating have proven difficult to identify. We find that the changes to the synthetic spectra over time are relatively small. If the features of the spectral profiles are caused by the non-thermal electrons, and these features are more or less sustained over the simulation duration, it should mean that small-scale events can be detected by instruments with slower cadence than the 1~s time step in this simulation as the signal remains relatively unchanged.
The spectral line diagnostics in this work include the \ion{Ca}{ii} lines, which gives potential for observing small-scale events with ground-based telescopes, such as the Swedish 1-m Solar Telescope \citep[SST;][]{2003SPIE.4853..341S}, 
the Daniel K. Inouye Solar Telescope \citep[DKIST;][]{2020SoPh..295..172R}, 
and the planned European Solar Telescope \citep[EST;][]{2022A&A...666A..21Q}. 
It is beneficial to include lines in the visible, as ground-based telescopes allow for higher spatial resolution compared to millimetre observations and extreme-UV diagnostics observed from space. Coordinated observations with for instance SST and IRIS would be advantageous to provide more constraints on small-scale heating events, even below the nanoflare limit. 

In this paper, we have investigated the effect of non-thermal electrons in a 3D Bifrost simulation by performing a detailed analysis of synthetic chromospheric and TR spectral lines. We have demonstrated that there is a clear difference between the spectra forming in regions subject to electron beams and not. We show that the spectral lines are highly affected by variations in vertical velocity and temperature, but the complexity of the atmospheric response in the Bifrost simulation makes it challenging to determine specific signatures that arise uniquely from the non-thermal electrons. Based on the simulations presented here, we cannot conclude that a clear and consistent signature will arise when higher beam energies are included. Additionally, the time span of the simulation is shorter than the typical lifetimes of small-scale heating events. A simulation with a longer time span and with higher energy beam heating events would be interesting to investigate when available. Still, the spectral line analysis performed in this work can contribute to the understanding of small-scale heating events in the solar atmosphere.

\begin{acknowledgements}
We thank Paola Testa for the useful comments that helped improve the paper.
This research was supported by the Research Council of Norway, project numbers 250810 and 
325491, 
and through its Centres of Excellence scheme, project number 262622. Computational resources have been provided by Sigma2 – the National Infrastructure for High-Performance Computing and Data Storage in Norway.
\end{acknowledgements}

\bibliographystyle{aa}
\bibliography{main.bib}
\end{document}